\documentclass{emulateapj}
\usepackage{amsmath}

\newcommand\kms{\ifmmode{\rm km\thinspace s^{-1}}\else km\thinspace s$^{-1}$\fi}
\newcommand\hip{{\it Hipparcos\/}}

\shortauthors{Torres et al.}
\shorttitle{Capella}

\begin{document}

\submitted{Accepted for publication in The Astrophysical Journal}

\title{Capella ($\alpha$\,Aurigae) revisited: new binary orbit,
  physical properties, and evolutionary state}

\author{
Guillermo Torres\altaffilmark{1},
Antonio Claret\altaffilmark{2},
Kre\v{s}imir Pavlovski\altaffilmark{3}, and
Aaron Dotter\altaffilmark{4}
}

\altaffiltext{1}{Harvard-Smithsonian Center for Astrophysics, 60
  Garden St., Cambridge, MA 02138, USA; e-mail:
  gtorres@cfa.harvard.edu}

\altaffiltext{2}{Instituto de Astrof\'\i sica de Andaluc\'\i a, CSIC,
  Apartado 3004, 18080 Granada, Spain; e-mail: claret@iaa.es}

\altaffiltext{3}{Department of Physics, Faculty of Science, University
  of Zagreb, Bijeni\u{c}ka cesta 32, 10000 Zagreb, Croatia; e-mail:
  pavlovski@phy.hr}

\altaffiltext{4}{Research School of Astronomy and Astrophysics,
  Australian National University, Canberra, ACT, 2611, Australia;
  e-mail: aaron.dotter@gmail.com}

\begin{abstract} 

Knowledge of the chemical composition and absolute masses of Capella
are key to understanding the evolutionary state of this benchmark
binary system comprising two giant stars. Previous efforts, including
our own 2009 study, have largely failed to reach an acceptable
agreement between the observations and current stellar evolution
models, preventing us from assessing the status of the primary. Here
we report a revision of the physical properties of the components
incorporating recently published high-precision radial velocity
measurements, and a new detailed chemical analysis providing
abundances for more than 20 elements in both stars. We obtain highly
precise ($\sim$0.3\%) masses of $2.5687\pm0.0074$\,$M_{\sun}$ and
$2.4828\pm0.0067$\,$M_{\sun}$, radii of $11.98\pm0.57$\,$R_{\sun}$ and
$8.83\pm0.33$\,$R_{\sun}$, effective temperatures of $4970\pm50$\,K
and $5730\pm60$\,K, and independently measured luminosities based on
the orbital parallax ($78.7\pm4.2$\,$L_{\sun}$ and
$72.7\pm3.6$\,$L_{\sun}$). We find an excellent match to stellar
evolution models at the measured composition of ${\rm [Fe/H]} =
-0.04\pm0.06$.  Three different sets of models place the primary star
firmly at the end of the core helium-burning phase (clump), while the
secondary is known to be evolving rapidly across the Hertzprung gap.
The measured lithium abundance, the C/N ratio, and the
$^{12}$C/$^{13}$C isotopic carbon abundance ratio, which change
rapidly in the giant phase, are broadly in agreement with expectations
from models.  Predictions from tidal theory for the spin rates,
spin-orbit alignment, and other properties do not fare as well,
requiring a 40-fold increase in the efficiency of the dissipation
mechanisms in order to match the observations.

\end{abstract}


\section{Introduction}
\label{sec:introduction}

As one of the brightest binary stars in the sky, Capella
($\alpha$\,Aurigae, HD\,34029, HR\,1708, HIP\,24608,
\ion{G8}{3}+\ion{G0}{3}, $P_{\rm orb} = 104$ days, $V = 0.07$) has
been studied for more than a century with a wide range of techniques
and at all observable wavelengths.\footnote{Capella has a wide common
  proper motion companion that is itself a visual binary composed of M
  dwarfs. The system is therefore a hierarchical quadruple. Revised
  properties of the M dwarf pair are reported in the Appendix.} A
persistent source of frustration for several decades has been the
difficulty in determining accurate absolute masses for the components,
despite the wealth of astrometric and spectroscopic measurements
available.  The history of this problem has been related by several
authors \citep[e.g.,][]{Batten:91, Barlow:93}, and most recently in
our earlier paper \citep[][hereafter T09]{Torres:09}. The challenge
associated with the masses has hindered efforts to pin down the
precise evolutionary state of the more massive primary star, which has
widely been considered to be a core helium-burning object, based
mostly on timescale arguments.  Disappointingly, current stellar
evolution models have so far largely failed to confirm that notion, as
it has not been possible to achieve a satisfactory fit to the global
properties of both stars simultaneously at a single age when assuming
the bulk chemical composition the system is believed to have.  The
secondary, on the other hand, is clearly on its way across the
Hertzprung gap.

The uncertainty in the masses was thought to have been solved in the
T09 analysis, which improved the formal precision by about a factor of
three compared to previous estimates, and documented efforts to
control systematic errors in the radial velocities that have likely
plagued the determination of the velocity amplitude of the
rapidly-rotating secondary star for decades, as described there.
Despite this, it was still not possible to establish the state of the
primary component unambiguously when enforcing a single age.

In the interim, \cite{Weber:11} have presented a new spectroscopic
study of Capella based on much higher-quality observational material
that leads to significantly larger masses for both stars than in our
2009 study, by several times the stated uncertainties. In particular,
the spread in mass between the stars increased from 1\% to about
3.5\%, which is an enormous difference for a pair of giants, and could
drastically change the assessment of their relative state of
evolution.  In addition, we have now made a new determination of the
chemical composition of Capella that is appreciably different from the
abundance assumed in the earlier paper, and is a key ingredient for
the comparison with stellar evolution models. These two developments
motivate us to take a fresh look at the system in order to investigate
the impact of the new measurements. Furthermore, \cite{Weber:11} have
presented evidence that the orbit of Capella may be very slightly
eccentric, unexpectedly, whereas all previous studies including our
own assumed it is circular. It is of interest, therefore, to revisit
our 2009 study of tidal evolution in the system (orbit circularization
and rotational synchronization) with more sophisticated models than
used there, especially in light of the new masses.

We have organized the paper as follows. In Sect.~\ref{sec:RVs} we
describe the new spectroscopic observations of \cite{Weber:11} and
comment on the issue of systematics in the radial velocities, in
comparison with our previous results from 2009. A revised orbital fit
for Capella is the presented in Sect.~\ref{sec:orbit}, using also
astrometric and other measurements from T09. Sect.~\ref{sec:chemical}
reports a new detailed chemical analysis of both stars from
disentangled spectra, along with a determination of the atmospheric
parameters. The revised physical properties of the stars are collected
in Sect.~\ref{sec:properties}, and are compared against three
different sets of stellar evolution models in
Sect.~\ref{sec:stellarevolution}. Key chemical diagnostics available
for Capella are also compared with model predictions in this section.
Then in Sect.~\ref{sec:tidalevolution} we examine the evolution of
orbital and stellar properties subjected to the influence of tidal
mechanisms, as a test of that theory. Sect.~\ref{sec:discussion}
presents a discussion of the results and concluding remarks. Finally,
the Appendix provides an update on the orbital properties of the wide
common proper motion companion of Capella.

\section{Radial velocities}
\label{sec:RVs}

The numerous historical radial-velocity (RV) measurements of Capella
have been discussed at length in our T09 study, which highlighted how
challenging it has been to determine accurate values for the rapidly
rotating secondary star ($v \sin i \approx 35$~\kms), whereas those of
the sharp-lined primary ($v \sin i \approx 4$~\kms) have been quite
consistent over the last century.  T09 presented 162 new RV
determinations for both components based on spectra obtained at the
Harvard-Smithsonian Center for Astrophysics (CfA) covering only a very
narrow wavelength range (45\,\AA). The RVs were measured using the
two-dimensional cross-correlation algorithm TODCOR \citep{Zucker:94},
with synthetic templates appropriate for each star.  Because of the
limited wavelength coverage, those measurements are susceptible to
systematic errors resulting mostly from lines shifting in and out of
the spectral window as a function of orbital phase.  Therefore, an
effort was made to control those biases by performing numerical
simulations to determine corrections to the velocities, which were at
the level of the final uncertainties in the individual measurements
for the secondary, and slightly larger for the primary.  Final errors
in the RVs as measured from the scatter in the orbital fit were about
0.44~\kms\ for the primary and 0.89~\kms\ for the secondary.  A sign
that systematic errors remained at some level in the CfA velocities
was evident in the residuals of the secondary star shown in Figure~2
of T09, in which a pattern can be seen as a function of orbital phase,
with a peak semi-amplitude of about twice the typical error.  Possible
explanations for this, as discussed by T09, include the presence of
spots on the active secondary star, or template mismatch.\footnote{In
  particular, due to limitations in the available library of synthetic
  spectra they used, the macroturbulent velocity of the templates
  ($\zeta_{\rm RT} = 1.5$~\kms) was not quite as large as appropriate
  for giant stars. This also resulted in an overestimation of the
  rotational velocities of the components, as discussed by T09.} An
additional indication of possible biases was the fact that a small
offset ($0.267 \pm 0.079$~\kms) was found between the primary and
secondary velocities in the global orbital fit of T09 that could not
be accounted for by differences in the gravitational redshift between
the stars, and was ascribed to similar reasons as the secondary
residual pattern.

More recently \cite{Weber:11} have reported new RV measurements for
Capella based on a very large set of more than 400 spectra obtained
with the STELLA \'echelle spectrograph \citep{Strassmeier:10} on a
1.2\,m robotic telescope in Tenerife, Spain. These spectra are of far
superior quality than those of T09, both in terms of wavelength
coverage (two orders of magnitude larger) and signal-to-noise
ratios. \cite{Weber:11} derived velocities using a similar
two-dimensional cross-correlation approach as T09, and also performed
numerical simulations to assess and correct for systematic errors
caused by the measuring technique. The velocity scatter from their
orbital fit is 0.064~\kms\ for the primary and 0.297~\kms\ for the
secondary, seven and three times smaller than in T09, respectively.
The key difference in this data set compared to T09 is in the
resulting velocity semi-amplitude of the secondary star ($K_{\rm B} =
26.840 \pm 0.024$~\kms), which is more than 6$\sigma$ larger than
reported by T09 ($K_{\rm B} = 26.260 \pm 0.087$~\kms). This difference
alone leads to absolute masses for Capella that are 4\% larger than 
in T09 for the primary, and 2\% for the secondary, a very significant
change that exceeds the formal mass errors by a factor of many.  The
semi-amplitudes $K_{\rm A}$ of the primary star, on the other hand,
are in virtually perfect agreement (see below).

Despite the much improved random errors of \cite{Weber:11}, the
residuals of the secondary velocities from their spectroscopic orbital
model (see their Figure~2) still display a phase-dependent pattern
reminiscent of the one in T09, also with a semi-amplitude of roughly
twice the errors, but at a much lower level in absolute terms.
Moreover, they note that a small offset is seen again between the
primary and secondary velocities (0.059~\kms) that cannot be explained
by the gravitational redshift effect.  This suggests that systematic
errors may still be lurking in these new measurements, for possibly
some of the same reasons as before. Nevertheless, any remaining
systematics are likely to be significantly less important than in T09,
as expected not only from the much higher quality of the spectroscopic
material, but suggested also by the significantly smaller magnitudes
of \emph{i}) the corrections for systematics applied by
\cite{Weber:11}, \emph{ii}) the formal uncertainties in the individual
RVs, or equivalently, the scatter from the spectroscopic orbital
solution, \emph{iii}) the amplitude of the residual pattern for the
secondary, and \emph{iv}) the unexplained residual primary/secondary
offset. In the next section we therefore incorporate these
measurements in a new orbital analysis of Capella.

Of the 438 RV measurements reported in Table~1 of \cite{Weber:11},
their final solution excluded 14 for the primary and 7 for the
secondary. We have done the same here as we found them to give
unusually large residuals, and we adopted also the measurement
uncertainties as published.

\section{Revised orbital fit}
\label{sec:orbit}

The global orbital solution in our T09 study of Capella combined all
usable astrometric observations in the literature with the CfA RVs for
both stars, as well as radial velocities for the primary star from
many of the historical data sets. The latter were carefully examined
to ensure that they imply $K_{\rm A}$ values consistent with those
from the CfA RVs in separate spectroscopic solutions using the same
fixed orbital period (see T09, Table~2). A similar solution of the
\cite{Weber:11} velocities shows that the primary semi-amplitude,
$K_{\rm A} = 29.959 \pm 0.005$~\kms, is essentially the same as that
from the CfA RVs, $K_{\rm A} = 29.96 \pm 0.04$~\kms. Therefore, for
our revised global solution below we have incorporated the
\cite{Weber:11} measurements for both stars, the CfA velocities for
the primary (but not the secondary), and the primary velocities from
the same historical data sets as used in T09. Although the
\cite{Weber:11} observations certainly dominate by weight, the older
measurements are still useful because they extend the baseline more
than a century, improving the orbital period.

The extensive astrometry available for Capella includes measurements
made by many authors with the technique of long-baseline
interferometry, beginning with the work of \cite{Merrill:22}, as well as
speckle interferometry, direct imaging, and the intermediate
astrometric measurements from \hip. These have all been described and
tabulated in our earlier study, and we refer the reader to that work
for details. So far as we are aware, no further astrometric
observations have been published for Capella except for those of
\cite{Huby:13}, which we do not use here, however, because of concerns
expressed by these authors about possible systematic errors
affecting their measurements.

Our global orbital fit follows closely that described by T09, and
includes the following parameters: orbital period ($P_{\rm orb}$),
relative angular semi-major axis ($a\arcsec$), inclination angle
($i_{\rm orb}$), eccentricity ($e$), longitude of periastron of the
more massive and cooler star ($\omega_{\rm A}$)\footnote{Note that
  this is the fainter star at optical wavelengths (see T09), which we
  refer to as star `A'.}, position angle of the ascending node for the
equinox J2000.0 ($\Omega_{\rm J2000}$), time of periastron passage
($T_{\rm peri}$), center-of-mass velocity ($\gamma$), the velocity
semi-amplitudes for each star ($K_{\rm A}$ and $K_{\rm B}$), the
angular semimajor axis of the photocenter ($a''_{\rm phot}$),
corrections to the \hip\ catalog values of the sky position of the
barycenter ($\Delta\alpha \cos\delta$, $\Delta\delta$) at the mean
catalog reference epoch of 1991.25, and corrections to the proper
motion components ($\Delta\mu_{\alpha} \cos\delta$,
$\Delta\mu_{\delta}$). To account for differences in the zero points
of the various RV data sets relative to the primary star measurements
of \cite{Weber:11}, we have also solved for 10 velocity offsets, one
for each set. An additional parameter, $f_\rho$, was included to
correct the scale of the angular separation measurements from two of
the astrometric data sets (see T09 for details).

\cite{Weber:11} discussed several adjustments made to their secondary
velocities to place them on the same system as their primary
velocities of Capella. These adjustments were intended to correct for
differences in the gravitational redshift of the two components, and
other shifts of unknown origin (see above).  From their discussion it
is not entirely clear to us whether these constant shifts have been
applied to the velocities they reported, so our global solution
includes one additional offset, $\Delta_{\rm AB}$, to account for
possible residual effects. With this, the total number of adjustable
parameters in our fit is 27.

The solution in our T09 study assumed a circular orbit for Capella, as
have all previous analyses of the binary. We noted, however, that
there were hints of a non-zero eccentricity in the interferometric
measurements of \cite{Hummel:94}, though not in the CfA RVs or in any
of the other data sets. We ascribed this to the transformation that
\cite{Hummel:94} made between their original interferometric
visibilities ($V^2$) and the nightly relative positions in polar coordinates
that they published, given that their own solution using the original
visibilities indicated a circular orbit. As pointed out by
\cite{Weber:11}, however, a spectroscopic fit using their RVs also
indicates a statistically significant non-zero eccentricity, of very
nearly the same magnitude as we had seen, and even with a consistent
longitude of periastron. This suggests we may have been too quick to
dismiss the possibility of a non-circular orbit in T09, as unexpected
as this may be for a pair of giants (one being a clump star) in a period of 104 days \citep[see, e.g.,][]{Massarotti:08}. We
discuss this further in Sect.~\ref{sec:tidalevolution}. For our new
global solution we have chosen to allow the orbit to be eccentric, on
the assumption that the effect is real.

\begin{deluxetable}{lc}
\tabletypesize{\scriptsize}
\tablecolumns{2}
\tablecaption{Revised orbital solution for Capella.\label{tab:elements}}
\tablehead{
\colhead{~~~~~~~~~~~~~~~Parameter~~~~~~~~~~~~~~~} & \colhead{Value}}
\startdata
\noalign{\vskip -4pt}
\sidehead{Adjusted quantities} \\
\noalign{\vskip -9pt}
~~~~$P_{\rm orb}$ (days)\dotfill               &  104.02128~$\pm$~0.00016\phn\phn \\
~~~~$a''$ (mas)\dotfill                        &  56.442~$\pm$~0.023\phn        \\
~~~~$i_{\rm orb}$ (deg)\dotfill                &  137.156~$\pm$~0.046\phn\phn  \\
~~~~$e$\dotfill                                &  0.00089~$\pm$~0.00011 \\
~~~~$\omega_{\rm A}$ (deg)\dotfill             &  342.6~$\pm$~9.0\phn\phn \\
~~~~$\Omega_{\rm J2000}$ (deg)\dotfill         &  40.522~$\pm$~0.039\phn       \\
~~~~$T_{\rm peri}$ (HJD$-$2,400,000)\dotfill   &  48147.6~$\pm$~2.6\phm{2222} \\
~~~~$\gamma$ (\kms)\tablenotemark{a}\dotfill   &  $+29.9387$~$\pm$~0.0032\phn\phs   \\
~~~~$K_{\rm A}$ (\kms)\dotfill                 &  25.9611~$\pm$~0.0044\phn       \\
~~~~$K_{\rm B}$ (\kms)\dotfill                 &  26.860~$\pm$~0.017\phn       \\
~~~~$a''_{\rm phot}$ (mas)\dotfill             &  2.14~$\pm$~0.70              \\
~~~~$\Delta\alpha^*$ (mas)\dotfill             &  $-$0.53~$\pm$~0.81\phs       \\
~~~~$\Delta\delta$ (mas)\dotfill               &  $-$0.37~$\pm$~0.57\phs       \\
~~~~$\Delta\mu_{\alpha} \cos\delta$ (mas~yr$^{-1}$)\dotfill     & $+$0.33~$\pm$~1.00\phs    \\
~~~~$\Delta\mu_{\delta}$ (mas~yr$^{-1}$)\dotfill                & $-$0.04~$\pm$~0.60\phs    \\
~~~~$f_{\rho}$\tablenotemark{b}\dotfill                         &  1.0400~$\pm$~0.0032          \\
~~~~$\Delta_{\rm AB}$ for WS (\kms)\tablenotemark{c}\dotfill    &  $+$0.050~$\pm$~0.013\phs        \\
~~~~$\Delta_{1\phn}$ $\langle$WS$-$C01$\rangle$ (\kms)\dotfill  &  $-$0.19~$\pm$~0.14\phs \\
~~~~$\Delta_{2\phn}$ $\langle$WS$-$N00$\rangle$ (\kms)\dotfill  &  $+$2.35~$\pm$~0.45\phs \\
~~~~$\Delta_{3\phn}$ $\langle$WS$-$G08$\rangle$ (\kms)\dotfill  &  $-$0.67~$\pm$~0.26\phs \\
~~~~$\Delta_{4\phn}$ $\langle$WS$-$S39$\rangle$ (\kms)\dotfill  &  $-$1.52~$\pm$~0.16\phs \\
~~~~$\Delta_{5\phn}$ $\langle$WS$-$S53$\rangle$ (\kms)\dotfill  &  $+$0.62~$\pm$~0.14\phs \\
~~~~$\Delta_{6\phn}$ $\langle$WS$-$B86$\rangle$ (\kms)\dotfill  &  $-$0.10~$\pm$~0.13\phs \\
~~~~$\Delta_{7\phn}$ $\langle$WS$-$S90$\rangle$ (\kms)\dotfill  &  $-$2.07~$\pm$~0.60\phs \\
~~~~$\Delta_{8\phn}$ $\langle$WS$-$B91$\rangle$ (\kms)\dotfill  &  $-$0.62~$\pm$~0.13\phs \\
~~~~$\Delta_{9\phn}$ $\langle$WS$-$B93$\rangle$ (\kms)\dotfill  &  $+$0.79~$\pm$~0.10\phs \\
~~~~$\Delta_{10}$ $\langle$WS$-$T09$\rangle$ (\kms)\dotfill     &  $+$0.289~$\pm$~0.035\phs \\
\sidehead{Derived quantities} \\
\noalign{\vskip -9pt}
~~~~$M_{\rm A}$ ($M_{\sun}$)\dotfill           &  2.5687~$\pm$~0.0074           \\
~~~~$M_{\rm B}$ ($M_{\sun}$)\dotfill           &  2.4828~$\pm$~0.0067          \\
~~~~$q\equiv M_{\rm B}/M_{\rm A}$\dotfill      &  0.96653~$\pm$~0.00062         \\
~~~~$a$ ($10^6$ km)\dotfill                    & 111.11~$\pm$~0.10\phn\phn     \\
~~~~$a$ ($R_{\sun}$)\dotfill                   & 159.72~$\pm$~0.15\phn\phn          \\
~~~~$a$ (au)\dotfill                           & 0.74272~$\pm$~0.00069          \\
~~~~$\pi_{\rm orb}$ (mas)\dotfill              &  75.994~$\pm$~0.089\phn         \\
~~~~Distance (pc)\dotfill                     &  13.159~$\pm$~0.015\phn       \\
~~~~$\mu_{\alpha} \cos\delta$ (mas~yr$^{-1}$)\dotfill    &  $+$75.85~$\pm$~1.00\phn\phs  \\
~~~~$\mu_{\delta}$ (mas~yr$^{-1}$)\dotfill               & $-$427.17~$\pm$~0.60\phn\phn\phs \\
~~~~$(\ell_{\rm B}/\ell_{\rm A})_{H_p}$\tablenotemark{d}\dotfill   & 1.204~$\pm$~0.060    
\enddata
\tablecomments{References for the RV offsets $\Delta_1$ to $\Delta_{10}$ are:
C01 = \cite{Campbell:01}; N00 = \cite{Newall:00}; G08 = \cite{Goos:08};
S39 = \cite{Struve:39}; S53 = \cite{Struve:53}; B86 = \cite{Beavers:86};
S90 = \cite{Shcherbakov:90}; B91 = \cite{Batten:91}; B93 = \cite{Barlow:93}; and
T09 = \cite{Torres:09}. The physical constants used in the analysis are those specified by \cite{Torresetal:10}.}
\tablenotetext{a}{On the reference frame of the RVs of \cite{Weber:11}.}
\tablenotetext{b}{Scale factor for the angular separation measurements by
\cite{Merrill:22} and \cite{Kulagin:70}.}
\tablenotetext{c}{Primary/secondary offset for the \cite{Weber:11} velocities (WS), in
the sense $\langle$primary \emph{minus} secondary$\rangle$.}
\tablenotetext{d}{Flux ratio between the secondary and primary in the \hip\ passband,
derived from the angular semimajor axis, the semimajor axis of the photocentric
orbit as measured by the satellite, and the velocity semi-amplitudes (see T09).}
\end{deluxetable}

The parameters of our fit are presented in Table~\ref{tab:elements},
along with other properties derived directly from the orbital
elements. With the exception of $K_{\rm B}$ and the quantities that
depend on it, the other results are rather similar to those in T09.
The eccentricity is essentially the same as derived by
\cite{Weber:11}.

\section{Spectroscopic analysis}
\label{sec:chemical}

Until recently the only detailed chemical analysis available for
Capella was that by \cite{McWilliam:90}, indicating a sub-solar
composition of ${\rm [Fe/H]} = -0.37 \pm 0.22$ on the scale of
\cite{Grevesse:84}, equivalent to ${\rm [Fe/H]} = -0.20 \pm 0.22$ on a
more modern scale in which the solar iron abundance is ${\rm A(Fe)} =
7.50$.\footnote{We use the standard abundance notation in which ${\rm
    A(X)} = \log [n({\rm X})/n({\rm H})]+12$, where $n({\rm X})$ and
  $n({\rm H})$ are the numbers of atoms per unit volume of element X
  and of hydrogen.}  This determination is presumably based on the
sharp lines of the primary star, but there is no indication that the
presence in the spectrum of the nearly equally bright secondary was
properly accounted for, and in addition, the analysis adopted an
incorrect primary temperature. A new metallicity determination was
reported by \cite{Fuhrmann:11} that gives a rather higher abundance of
${\rm [Fe/H]} = +0.05 \pm 0.08$, apparently on the scale of
\cite{Grevesse:96} in which the solar iron abundance is also ${\rm
  A(Fe)} = 7.50$. This study is based on spectral synthesis applied to
eight \ion{Fe}{1} lines and one \ion{Fe}{2} line from a single
composite spectrum with some degree of line blending, using an
unspecified primary/secondary flux ratio.

Below we describe our new determination of the chemical composition
and atmospheric parameters of Capella based on the technique of
spectral disentangling \citep{Simon:94}, which bypasses the line
blending problems inherent in previous analyses that used composite
spectra.

\subsection{Disentangling}
\label{sec:disentangling}

\begin{figure}
\epsscale{1.15}
\plotone{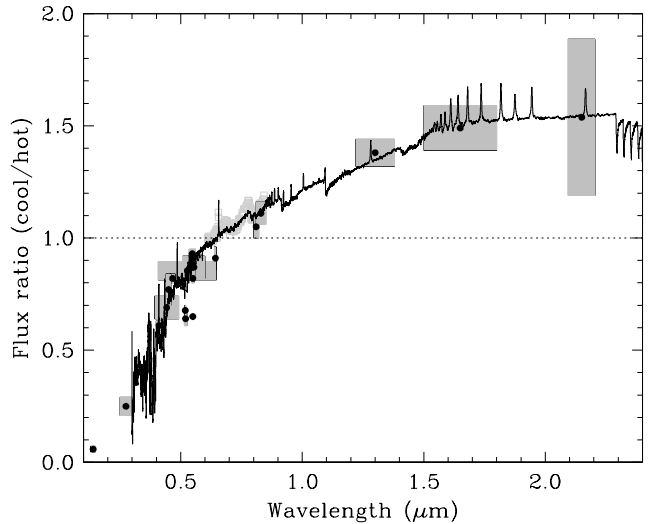}

\figcaption[]{Measured flux ratios for the components of Capella (cool
  star relative to hot star, i.e., primary relative to
  secondary). Values from T09 are indicated with dots and error boxes,
  in which the horizontal length of each box indicates the wavelength
  coverage. Other measurements from \cite{Huby:13} are shown by the
  lighter gray squares in the range 0.611--0.843 $\mu$m. Overplotted is
  the predicted flux ratio based on synthetic spectra by
  \cite{Husser:13} scaled according to the radius ratio given by
  T09.\label{fig:fluxratio}}

\end{figure}

\begin{figure}
\epsscale{1.15}
\plotone{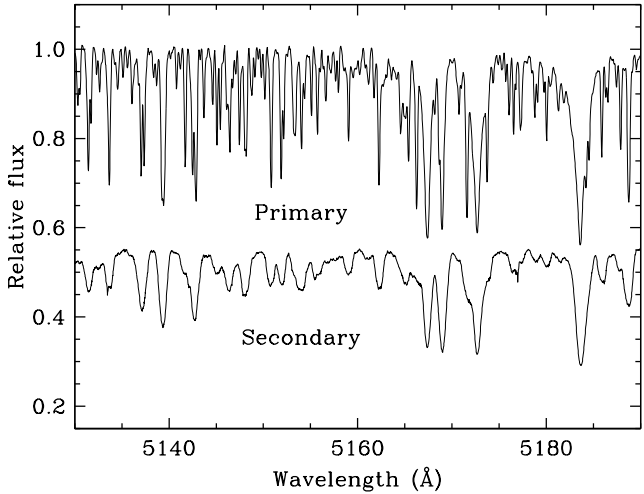}

\figcaption[]{Portions of the disentangled ELODIE spectra of
  Capella. Secondary shifted vertically for
  clarity.\label{fig:disentangled}}

\end{figure}

For our spectroscopic analysis we made use of public archival spectra
taken in 2003 and 2004 with the ELODIE spectrograph \citep{Baranne:96}
on the 1.93\,m telescope at the Observatoire de Haute-Provence, in
France. The nominal resolving power of the instrument is $R = 42,000$,
and the 15 spectra used span the approximate wavelength range
4000--6800\,\AA, with signal-to-noise ratios ranging from about 130 to
560 per pixel at 5550\,\AA. We have disentangled these spectra using
the {\sc FDBinary} program of \cite{Ilijic:04}, in the same way as
described recently by \cite{Torres:14}. {\sc FDBinary} implements
spectral disentangling in the Fourier domain according to
\cite{Hadrava:95}. The signal-to-noise ratios of the resulting
disentangled spectra are approximately 510 for the cooler primary and
590 for the secondary. Renormalization of the disentangled spectra for
a proper abundance analysis \citep[see][]{Pavlovski:05, Lehmann:13}
requires knowledge of the relative flux contribution of each star at
each wavelength.  Flux ratios for Capella have been measured
throughout the UV, optical, and near-infrared range, as reported by
T09.  Figure~\ref{fig:fluxratio} shows the predicted flux ratio based
on PHOENIX model spectra from \cite{Husser:13} for parameters near
those of the components, along with the measurements tabulated by T09
as well as others from the recent study by \cite{Huby:13} over the
range 6112--8430\,\AA. The agreement is very good. For our purposes we
have used a smoothed version of this relation. Sample segments of the
disentangled spectra of the two components of Capella are presented in
Figure~\ref{fig:disentangled}.

\subsection{Atmospheric parameters and abundance analysis}
\label{sec:abundances}

The general methodology for determining the atmospheric parameters and
abundances from the disentangled spectra follows the procedures
described by \cite{Torres:14}. The {\sc uclsyn} code
\citep{Smalley:01} was used to synthesize spectra, based on initial
temperatures and surface gravities from T09 and a built-in grid of LTE
model atmospheres by \cite{Castelli:03} with a scaled-solar mixture. Line broadening was modeled
adopting an initial composition matching the Sun, and microturbulent
velocities of $\xi_{\rm t} = 1.5$\,\kms\ for both components
(fine-tuned below).  Least-squares fitting then yielded a
macroturbulent velocity for the primary of $\zeta_{\rm A} = 6.6 \pm
0.4$\,\kms, and a projected rotational velocity of $v_{\rm A} \sin i =
4.4 \pm 0.5$\,\kms.  Given the rapid rotation of the secondary,
macroturbulence has a negligible effect on the line profiles and we
made no attempt to determine it from the ELODIE spectra. The
rotational velocity of the companion was measured to be $v_{\rm B}
\sin i = 34.5 \pm 0.7$\,\kms.

Elemental abundances were determined using spectral lines suitable for
giant stars over the wavelength range 4600--6750\,\AA. Line lists and
atomic data were taken from the work of \cite{Reddy:12},
\cite{Bocek:15}, and \cite{Lyubimkov:15}.  Equivalent widths measured
within {\sc uclsyn} for the very numerous iron lines were used to set
the microturbulent velocities from the condition of a null correlation
between the abundance and the reduced equivalent widths. We derived
values of $\xi_{\rm t,A} = 1.48 \pm 0.08$\,\kms\ for the primary and
$\xi_{\rm t,B} = 1.55 \pm 0.11$\,\kms\ for the secondary. We also made
an estimate of the effective temperatures from the usual condition of
excitation equilibrium, iterating with the measurement of $\xi_{\rm
  t}$, with the following results: $T_{\rm eff,A} = 4980 \pm 80$\,K
and $T_{\rm eff,B} = 5750 \pm 110$\,K. There is very good agreement
between these values and others reported by T09; we discuss them
further below. The surface gravities in our analysis were held fixed
at the estimates reported by T09, which are very close to our final
values described in the next section.

Detailed abundances were determined for 22 species in both stars, and
oxygen in the primary only. They are listed in
Table~\ref{tab:abundances}, which includes also the values relative to
the Sun on the scale of \cite{Asplund:09}. No adjustments have been
applied for NLTE effects. The uncertainties account for possible
errors in $T_{\rm eff}$ as reported above, and also include a
contribution from a representative error of 0.1\,\kms\ in $\xi_{\rm
  t}$. The uncertainties in $\log g$ have a negligible impact. 
  The choice of the mixture adopted in the model atmospheres,
  particularly the CNO composition, also has a minimal effect on our
  abundance determinations \citep[see also][]{Morel:14}. We find no
  dependence of the Fe abundance on wavelength, which is an indication
  that our adopted wavelength-dependent flux ratios (Figure 1) are
  accurate and do not introduce significant systematic errors in the
  abundances. A further indication of the robustness of our
  determinations is the fact that the abundances are quite similar for
  the two components (except for species affected by evolution; see
  below), as expected for a binary system.

\begin{figure}
\epsscale{1.15}
\plotone{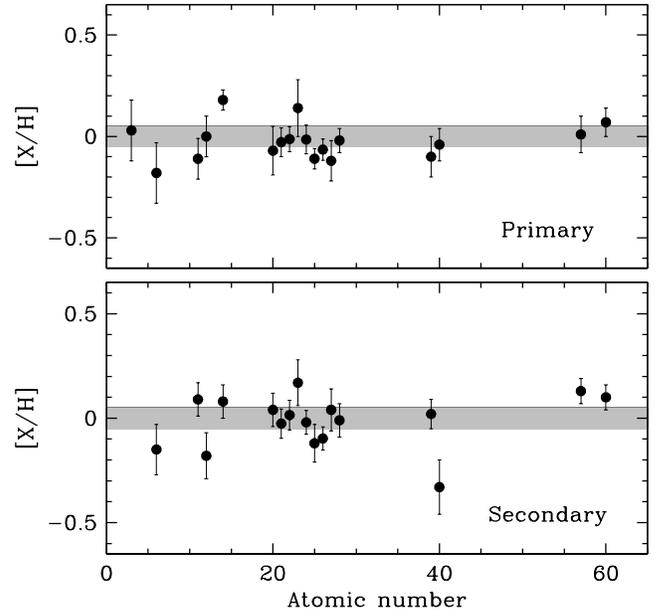}

\figcaption[]{Photospheric abundance pattern measured for the Capella
  components, compared to the standard solar composition of
  \cite{Asplund:09} (gray shading). Abundances from different ions of
  the same element have been averaged.\label{fig:abundances}}

\end{figure}

The weighted average iron abundance of the two stars from \ion{Fe}{1}
is ${\rm [Fe/H]} = -0.04 \pm 0.06$, or very nearly solar, in contrast
with the sub-solar composition adopted by T09, and in better agreement
with the estimate by \cite{Fuhrmann:11}.  We find no significant
enhancement of the $\alpha$ elements in Capella: ${\rm [\alpha/Fe]} =
-0.02 \pm 0.04$. A graphical representation of the abundance pattern
for the two components is seen in Figure~\ref{fig:abundances},
compared to the solar composition.

The lithium abundance has long been known to be very different for the
two components of this binary \citep{Wallerstein:64, Wallerstein:66} as
a result of chemical evolution in the primary. We used spectral lines
and atomic data from \cite{Lyubimkov:12} in the vicinity of the
\ion{Li}{1} $\lambda$6708 doublet to make new estimates for each star,
and obtained values of ${\rm A(Li)} = 1.08 \pm 0.11$ and ${\rm A(Li)}
= 3.28 \pm 0.13$ for the primary and secondary, respectively. These
are consistent with previous measurements. The equivalent widths we
determined are $21.4 \pm 0.7$\,m\AA\ and $297.4 \pm 8.0$\,m\AA.

\begin{deluxetable*}{ll c ccc c ccc c c}
\tablewidth{0pc}
\tablecaption{Abundances from our disentangled ELODIE spectra of Capella.\label{tab:abundances}}
\tablehead{
\colhead{} &
\colhead{} & &
\multicolumn{3}{c}{Primary} & &
\multicolumn{3}{c}{Secondary} & &
\colhead{} \\
\cline{4-6} \cline{8-10} \\ [-1ex]
\colhead{A} &
\colhead{X} & &
\colhead{Abundance} &
\colhead{[X/H]} &
\colhead{$N$} &&
\colhead{Abundance} &
\colhead{[X/H]} & 
\colhead{$N$} & &
\colhead{$\log\epsilon_{\sun}$}
}
\startdata
  3 & \ion{Li}{1}  &&   $1.08 \pm 0.11$ &  $-0.07 \pm 0.15$ &  3  &&    $3.28 \pm 0.13$ &  $+2.30 \pm 0.16$  &  3 &&  $1.05 \pm 0.10$ \\
  6 & \ion{C}{1}   &&   $8.25 \pm 0.14$ &  $-0.18 \pm 0.15$ &  4  &&    $8.28 \pm 0.11$ &  $-0.15 \pm 0.12$  &  5 &&  $8.43 \pm 0.05$ \\
  8 & \ion{O}{1}   &&   $8.55 \pm 0.11$ &  $-0.14 \pm 0.12$ &  1  &&       \nodata      &     \nodata        &  \nodata  &&  $8.69 \pm 0.05$ \\
 11 & \ion{Na}{1}  &&   $6.13 \pm 0.09$ &  $-0.11 \pm 0.10$ &  4  &&    $6.33 \pm 0.07$ &  $+0.09 \pm 0.08$  &  4 &&  $6.24 \pm 0.04$ \\
 12 & \ion{Mg}{1}  &&   $7.60 \pm 0.09$ &  $+0.00 \pm 0.10$ &  2  &&    $7.42 \pm 0.10$ &  $-0.18 \pm 0.11$  &  5 &&  $7.60 \pm 0.04$ \\
 14 & \ion{Si}{1}  &&   $7.69 \pm 0.04$ &  $+0.18 \pm 0.05$ & 10  &&    $7.59 \pm 0.07$ &  $+0.08 \pm 0.08$  &  7 &&  $7.51 \pm 0.03$ \\
 20 & \ion{Ca}{1}  &&   $6.27 \pm 0.11$ &  $-0.07 \pm 0.12$ &  8  &&    $6.38 \pm 0.11$ &  $+0.04 \pm 0.08$  &  7 &&  $6.34 \pm 0.04$ \\
 21 & \ion{Sc}{1}  &&   $3.13 \pm 0.15$ &  $-0.02 \pm 0.16$ &  5  &&    $3.16 \pm 0.10$ &  $+0.01 \pm 0.11$  &  6 &&  $3.15 \pm 0.04$ \\
 21 & \ion{Sc}{2}  &&   $3.12 \pm 0.07$ &  $-0.03 \pm 0.08$ &  8  &&    $3.10 \pm 0.08$ &  $-0.05 \pm 0.09$  & 11 &&  $3.15 \pm 0.04$ \\
 22 & \ion{Ti}{1}  &&   $4.96 \pm 0.12$ &  $+0.01 \pm 0.13$ & 15  &&    $5.02 \pm 0.09$ &  $+0.07 \pm 0.10$  & 11 &&  $4.95 \pm 0.05$ \\
 22 & \ion{Ti}{2}  &&   $4.93 \pm 0.05$ &  $-0.02 \pm 0.07$ &  3  &&    $4.91 \pm 0.09$ &  $-0.04 \pm 0.10$  &  6 &&  $4.95 \pm 0.05$ \\
 23 & \ion{V}{1}   &&   $4.07 \pm 0.12$ &  $+0.14 \pm 0.14$ & 14  &&    $4.10 \pm 0.07$ &  $+0.17 \pm 0.11$  & 13 &&  $3.93 \pm 0.08$ \\
 24 & \ion{Cr}{1}  &&   $5.64 \pm 0.09$ &  $+0.00 \pm 0.10$ &  9  &&    $5.67 \pm 0.07$ &  $+0.03 \pm 0.08$  & 11 &&  $5.64 \pm 0.04$ \\
 24 & \ion{Cr}{2}  &&   $5.61 \pm 0.09$ &  $-0.03 \pm 0.10$ &  7  &&    $5.57 \pm 0.07$ &  $-0.07 \pm 0.08$  &  6 &&  $5.64 \pm 0.04$ \\
 25 & \ion{Mn}{1}  &&   $5.32 \pm 0.09$ &  $-0.11 \pm 0.05$ &  8  &&    $5.31 \pm 0.08$ &  $-0.12 \pm 0.09$  &  5 &&  $5.43 \pm 0.05$ \\
 26 & \ion{Fe}{1}  &&   $7.47 \pm 0.06$ &  $-0.03 \pm 0.07$ & 42  &&    $7.44 \pm 0.08$ &  $-0.06 \pm 0.09$  & 41 &&  $7.50 \pm 0.04$ \\
 26 & \ion{Fe}{2}  &&   $7.39 \pm 0.07$ &  $-0.11 \pm 0.08$ &  8  &&    $7.38 \pm 0.06$ &  $-0.12 \pm 0.07$  & 11 &&  $7.50 \pm 0.04$ \\
 27 & \ion{Co}{1}  &&   $4.87 \pm 0.08$ &  $-0.12 \pm 0.10$ &  8  &&    $5.03 \pm 0.07$ &  $+0.04 \pm 0.10$  &  5 &&  $4.99 \pm 0.07$ \\
 28 & \ion{Ni}{1}  &&   $6.20 \pm 0.04$ &  $-0.02 \pm 0.06$ & 16  &&    $6.21 \pm 0.07$ &  $-0.01 \pm 0.08$  & 17 &&  $6.22 \pm 0.04$ \\
 39 & \ion{Y}{2}   &&   $2.11 \pm 0.09$ &  $-0.10 \pm 0.10$ &  4  &&    $2.23 \pm 0.05$ &  $+0.02 \pm 0.07$  &  5 &&  $2.21 \pm 0.05$ \\
 40 & \ion{Zr}{1}  &&   $2.54 \pm 0.07$ &  $-0.04 \pm 0.08$ &  5  &&    $2.25 \pm 0.12$ &  $-0.33 \pm 0.13$  &  3 &&  $2.58 \pm 0.04$ \\
 57 & \ion{La}{2}  &&   $1.11 \pm 0.08$ &  $+0.01 \pm 0.09$ &  5  &&    $1.23 \pm 0.05$ &  $+0.13 \pm 0.06$  &  5 &&  $1.10 \pm 0.04$ \\
 60 & \ion{Nd}{2}  &&   $1.49 \pm 0.06$ &  $+0.07 \pm 0.07$ &  8  &&    $1.52 \pm 0.05$ &  $+0.10 \pm 0.06$  &  7 &&  $1.42 \pm 0.04$ 
\enddata

\tablecomments{Columns list the atomic number, the element and
  ionization degree, the logarithm of the number abundance on the
  usual scale in which A(H) = 12, the logarithmic abundance relative
  to the Sun, and the number of spectral lines measured.  The last
  column gives the reference photospheric solar values from
  \cite{Asplund:09}.  }

\end{deluxetable*}

Additional chemical diagnostics for Capella have been reported by T09,
and include the $^{12}$C/$^{13}$C carbon isotope ratio for the primary
star \citep[$27 \pm 4$;][]{Tomkin:76} and the C/N abundance ratios for
both components ($0.57 \pm 0.06$ for the primary and $3.30 \pm 0.16$
for the secondary). We adopt those as published.

\section{Physical properties}
\label{sec:properties}

The revised masses of Capella that incorporate the new RVs of
\cite{Weber:11} have formal uncertainties of 0.3\%, and differ only
slightly from theirs through a combination of a marginally larger
$K_{\rm B}$ value in our analysis and a smaller orbital inclination
angle than they used. The new masses are 4.2\% and 1.6\% larger than
those listed by T09, which is a significant difference due almost
entirely to the change in the velocity amplitude of the secondary. The
radii, based on the individual angular diameters from T09 and the
revised orbital parallax, are approximately 0.9\% larger than before,
and have precisions of about 5\% and 4\% for the primary and
secondary, respectively. These are limited by the angular diameters,
as the uncertainty in the orbital parallax is only 0.11\%.

\begin{deluxetable}{lc@{~~~~}c}
\tablecolumns{3}
\tablewidth{0pc}
\tablecaption{Revised physical parameters of Capella.\label{tab:dimensions}}
\tablehead{
\colhead{~~~~~~~~Parameter~~~~~~~~} & 
\colhead{Primary} &
\colhead{Secondary}
}
\startdata
Mass ($M_{\sun}$)\dotfill              &  2.5687~$\pm$~0.0074 & 2.4828~$\pm$~0.0067  \\
$q\equiv M_{\rm B}/M_{\rm A}$\dotfill  & \multicolumn{2}{c}{0.96653~$\pm$~0.00062} \\
$a$ ($10^6$ km)\dotfill                & \multicolumn{2}{c}{111.11~$\pm$~0.10\phn\phn}  \\
$a$ (au)\dotfill                       & \multicolumn{2}{c}{0.74272~$\pm$~0.00069}   \\
$\pi_{\rm orb}$ (mas)\dotfill          & \multicolumn{2}{c}{75.994~$\pm$~0.089\phn}  \\
Distance (pc)\dotfill                  & \multicolumn{2}{c}{13.159~$\pm$~0.015\phn}   \\
Radius ($R_{\sun}$)\dotfill            & 11.98~$\pm$~0.57\phn & 8.83~$\pm$~0.33 \\
$\log g$ (cgs)\dotfill                 & 2.691~$\pm$~0.041 & 2.941~$\pm$~0.032 \\
$T_{\rm eff}$ (K)\dotfill              & 4970~$\pm$~50\phn\phn & 5730~$\pm$~60\phn\phn \\
Luminosity ($L_{\sun}$)\tablenotemark{a}\dotfill        & 78.7~$\pm$~4.2\phn & 72.7~$\pm$~3.6\phn \\
$BC_V$ (mag)\dotfill                     & $-$0.304~$\pm$~0.055\phs & $-$0.089~$\pm$~0.051\phs \\
$M_V$ (mag)\dotfill                    & 0.296~$\pm$~0.016 & 0.167~$\pm$~0.015 \\
$v \sin i$ (\kms)\tablenotemark{b}\dotfill              & 4.1~$\pm$~0.4 & 35.0~$\pm$~0.5\phn \\
$P_{\rm rot}$ (days)\tablenotemark{c}\dotfill  & 104~$\pm$~3\phn\phn & 8.5~$\pm$~0.2 \\
Age (Myr)\tablenotemark{d}\dotfill                      & \multicolumn{2}{c}{590--650} \\
${\rm [Fe/H]}$\dotfill                         & \multicolumn{2}{c}{$-$0.04~$\pm$~0.06\phs} \\
A(Li)\dotfill                          & 1.08~$\pm$~0.11 & 3.28~$\pm$~0.13 \\
$^{12}$C/$^{13}$C\,\tablenotemark{d}\dotfill              & 27~$\pm$~4\phn & \nodata \\
C/N\,\tablenotemark{c}\dotfill           & 0.57~$\pm$~0.06  & 3.30~$\pm$~0.16 
\enddata
\tablenotetext{a}{Computed from $V$, $\pi_{\rm orb}$, and $BC_V$ from
\cite{Flower:96}, adopting $M_{\rm bol}^{\sun} = 4.732$ (see T09 and \citealt{Torres:10}).}
\tablenotetext{b}{Average of 5 measurements from the literature for the primary and 10 for
  the secondary that account for macroturbulence, including our own
  (see text).}
\tablenotetext{c}{Measured values adopted from T09.}
\tablenotetext{d}{Age range from the MESA and Granada models (see text).}
\tablenotetext{e}{Measurement by \cite{Tomkin:76}.}
\end{deluxetable}

T09 reported three independent estimates of the effective temperatures
for the two components. One is from a comparison of their
spectroscopic observations with synthetic spectra with solar
metallicity, giving $T_{\rm eff,A} = 4900 \pm 100$\,K and $T_{\rm
  eff,B} = 5710 \pm 100$\,K. Another came from the use of the measured
angular diameters of the stars along with their apparent magnitudes,
the parallax, and bolometric corrections. The updated parallax in the
present work does not alter those estimates significantly; they are
$T_{\rm eff,A} = 4970 \pm 160$\,K and $T_{\rm eff,B} = 5690 \pm
130$\,K. A third determination by T09 was based on the measured color
indices for the stars, and the use of the color/temperature
calibrations of \cite{Ramirez:05}, which depend on metallicity. A
sub-solar composition ${\rm [m/H]} = -0.37 \pm 0.07$ had been assumed
by T09, whereas we now derive a value much closer to solar. Using our
determination of ${\rm [Fe/H]} = -0.04 \pm 0.06$
(Sect.~\ref{sec:abundances}), the revised photometric estimates become
4940\,K and 5680\,K, which are 30\,K and 70\,K higher than
before. Furthermore, a careful examination of the zero point of the
\cite{Ramirez:05} calibrations by \cite{Casagrande:10} suggests that
the scale of those relations is too cool by about 85\,K compared to
the best available absolute scale, at least in the temperature range
of the Capella components. We have therefore applied this offset,
obtaining corrected photometric estimates of $T_{\rm eff,A} = 5025 \pm
110$\,K and $T_{\rm eff,B} = 5765 \pm 120$\,K. The uncertainties
include a contribution of 100\,K added in quadrature to the
photometric and calibration errors, to be conservative. Finally, a
fourth temperature determination was reported in the previous section
from the disentangled ELODIE spectra, giving $T_{\rm eff,A} = 4980 \pm
80$\,K and $T_{\rm eff,B} = 5750 \pm 110$\,K. The weighted average of
the four values for each component is $T_{\rm eff,A} = 4970 \pm 50$\,K
and $T_{\rm eff,B} = 5730 \pm 60$\,K, in which the uncertainties
account not only for the individual weights but also for the scatter
of the measurements, and are believed to be realistic. These averages
are 50\,K hotter than in T09.

The available determinations of $v \sin i$ for both components were
summarized in our previous work (T09, Table~14). Our present
measurements from the ELODIE spectra are consistent with those of
others, as well as with the measurements reported by
\cite{Fuhrmann:11}, which are $v_{\rm A} \sin i = 3.5 \pm
0.8$\,\kms\ and $v_{\rm B}\sin i = 35.4 \pm 3.2$\,\kms. The weighted
averages of all independent determinations (5 for the primary, 10 for
the secondary) that have taken account of macroturbulence broadening,
especially for the primary component, are $v_{\rm A}\sin i = 4.1 \pm
0.4$\,\kms\ and $v_{\rm B}\sin i = 35.0 \pm 0.5$\,\kms, which we adopt
for the remainder of the paper.

The masses, radii, temperatures, and other derived properties are
summarized in Table~\ref{tab:dimensions}. Note that the bolometric
luminosities are independent of the temperatures and radii, and are
based on the apparent magnitudes, the orbital parallax, and bolometric
corrections from \cite{Flower:96}, as in T09. If we instead compute
them from $T_{\rm eff}$ and $R$, the results are consistent, but have
larger formal uncertainties: $L_{\rm A} = 78.7 \pm 8.1$\,$L_{\sun}$
and $L_{\rm B} = 75.4 \pm 6.4$\,$L_{\sun}$.

\section{Comparison with stellar evolution models}
\label{sec:stellarevolution}

Up until now the ability of stellar evolution models to match all of
the global properties of both components of Capella simultaneously at
a single age has not been entirely satisfactory, likely at least in
part because there are so many observational constraints
available. This has made it difficult to establish the evolutionary
status of the primary star unambiguously, although it has widely been
thought to be a core helium-burning (clump) star, based on timescale
arguments (see T09).  The significantly different (and more precise)
masses obtained above, and evidence that the chemical composition is
rather different from that previously assumed, motivate us to revisit
the comparison with stellar evolution models here.

An initial test was made using the PARSEC isochrones of
\cite{Bressan:12}.\footnote{http://stev.oapd.inaf.it/cmd~.} These
models adopt the solar distribution of heavy elements from the
compilation by \cite{Grevesse:98}, with adjustments to some elements
following \cite{Caffau:11} such that the solar photospheric
metallicity is $Z_{\sun} = 0.01524$. On this scale the measured
abundance of Capella (Sect.~\ref{sec:abundances}) corresponds
approximately to $Z = 0.0133$. The helium abundance $Y$ follows an
adopted enrichment law with a slope $\Delta Y/\Delta Z = 1.78$, which
results in a value for Capella of $Y = 0.272$. Convection is treated
in the standard mixing length theory approximation. The calibration to
the Sun leads to a mixing length parameter of $\alpha_{\rm MLT} =
1.74$, which is held fixed in these models. Convective core
overshooting is also included, with an efficiency parameter of
$\Lambda_{\rm c} = 0.5$ for Capella, representing the mean free path
of convective bubbles across the border of the convective region,
expressed in units of the pressure scale height $H_{\rm p}$.  This is
roughly equivalent to $\alpha_{\rm ov} = 0.25$ pressure scale heights
above the convective boundary in the more commonly used formulation of
this phenomenon (see below).  Mass loss from stellar winds is not
considered in these models for the mass range of interest for Capella,
nor is rotation.

\begin{figure}
\epsscale{1.15}
\plotone{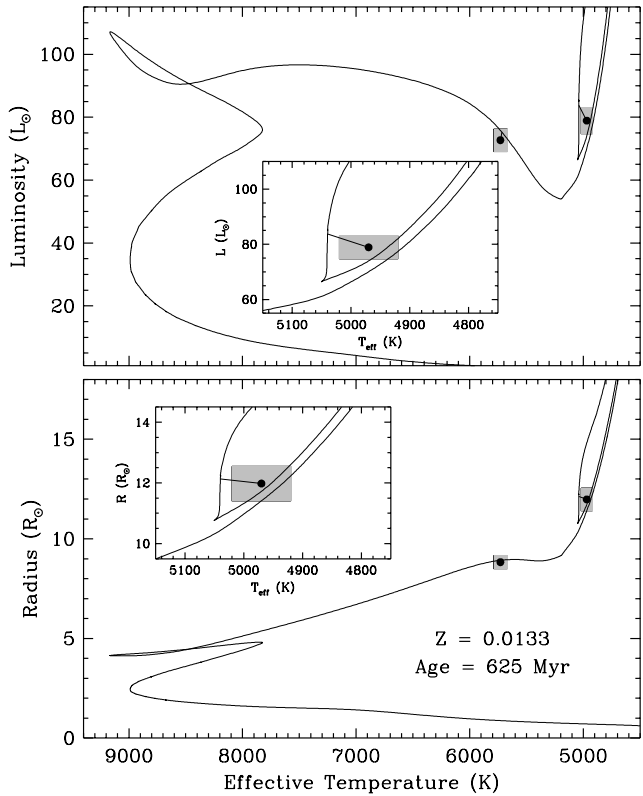}

\figcaption[]{Comparison of the observed properties of Capella ($M$,
  $R$, $T_{\rm eff}$, $L$, $q$) with a PARSEC isochrone from the model
  series by \cite{Bressan:12} in the $L$--$T_{\rm eff}$ and
  $R$--$T_{\rm eff}$ planes. Metallicity and age are as indicated in
  the lower panel. The insets show enlargements around the primary
  star, with a short line segment connecting the measured location in
  each plane with the best-fit position along the isochrone.
  Corresponding insets for the secondary are not shown as the match to
  its properties is better.\label{fig:girardi}}

\end{figure}

With the chemical composition and convective parameters fixed as
described above, we searched for the common age giving the best fit to
the masses, radii, temperatures, and luminosities of both stars using
the $\chi^2$ statistic
\begin{displaymath}
\chi^2 = \sum\left(\left[\frac{\Delta M}{\sigma_M}\right]^2 +
\left[\frac{\Delta T_{\rm eff}}{\sigma_{T_{\rm eff}}}\right]^2 +
\left[\frac{\Delta L}{\sigma_L}\right]^2 +
\left[\frac{\Delta R}{\sigma_R}\right]^2\right)
\end{displaymath}
as the figure of merit, where the sum is over both stars and the
$\Delta$ quantities represent the difference between the predicted and
measured properties. A reasonably good fit was obtained for an age of
622 Myr, matching the properties of the stars within about 1.4 times
their uncertainties, with the largest discrepancy being in the primary
temperature. We note, however, that the mass ratio of Capella is known
much more precisely than the individual masses, with $\sigma_q \approx
0.06$\% compared to mass errors $\sigma_M$ of 0.29\% and 0.27\%.  The
masses in the above fit were allowed to vary independently, and as a
result the best-fit mass ratio differs from the measured value by
about 3$\sigma$. We therefore repeated the fit constraining $q$ to be
near its measured value by using a corresponding penalty term $[\Delta
  q/\sigma_q]^2$ in $\chi^2$ instead of the secondary mass term.  We
obtained a solution of similar quality (all properties reproduced
within 1.4$\sigma$) and nearly the same age of 625 Myr.  This fit is
illustrated in Figure~\ref{fig:girardi}, and it places the primary
star in the core helium-burning phase (clump).

Aside from the global structural properties of the stars considered
above, several chemical diagnostics including the lithium abundance,
the C/N ratios, and the isotopic carbon abundance ratio
$^{12}$C/$^{13}$C are available for Capella that are not normally
tabulated with published stellar evolution models, but that are
nevertheless interesting to compare with theoretical predictions. To
that end, we have performed an additional test against a second set of
models by \cite{Claret:04}, occasionally referred to below as the
Granada models.

These models adopt also the solar abundance distribution by
\cite{Grevesse:98}, with some adjustments such that the solar
metallicity becomes $Z_{\sun} = 0.0189$. The abundance of Capella then
corresponds approximately to $Z = 0.0172$, which we held fixed. The
enrichment law $\Delta Y/\Delta Z = 2.0$ typically used in these
models results in a helium abundance for Capella of $Y = 0.274$,
similar to that used above.  The solar-calibrated value of the mixing
length parameter is $\alpha_{\rm MLT} = 1.68$, and convective core
overshooting $\alpha_{\rm ov}$ is parametrized such that the mean free
path above the convective boundary is $d_{\rm ov} = \alpha_{\rm ov}
H_{\rm p}$. Rotation was initially not included in our tests.

\begin{figure}[t]
\epsscale{1.15}
\plotone{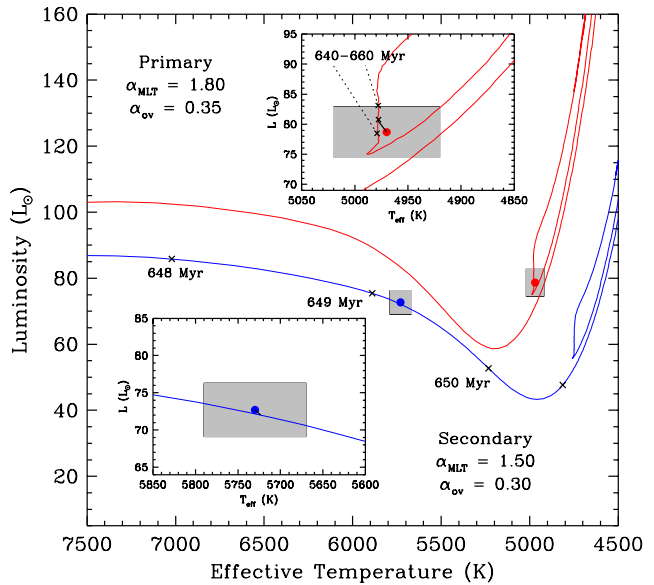}

\figcaption[]{Comparison of the observed properties of Capella (dots
  and error boxes) with the Granada models by \cite{Claret:04} in the
  $L$--$T_{\rm eff}$ diagram. The evolutionary tracks shown are for
  the measured masses (incremented by 0.005\,$M_{\sun}$ for the
  primary and 0.002\,$M_{\sun}$ for the secondary, to account for mass
  loss; see text) and the measured metallicity ($Z = 0.0172$ for these
  models). Reference ages are marked along the secondary track, and
  the insets show enlargements around the position of each star, with
  a short line segment connecting the observations to the predicted
  positions.\label{fig:claret1}}

\end{figure}

\begin{figure}
\epsscale{1.15}
\plotone{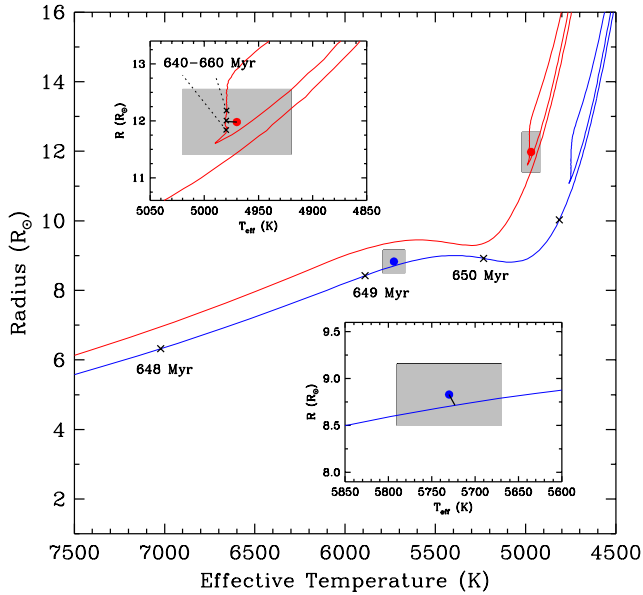}

\figcaption[]{Similar to Figure~\ref{fig:claret1} for the $R$--$T_{\rm
    eff}$ plane.\label{fig:claret2}}

\end{figure}

A grid of Granada evolutionary tracks was computed for the measured
masses and a range of convective parameters for each component of
Capella, as in principle there is no reason to expect stars in such
different evolutionary states to have the same convective properties.
We varied $\alpha_{\rm MLT}$ between values of 1.0 and 2.2 in steps of
0.1, and $\alpha_{\rm ov}$ over the range 0.15--0.40, with a step of
0.05.  Mass loss was included in these calculations following the
prescription by \cite{Reimers:75}.  Preliminary tests indicated a
minimal loss of mass for both stars up to the present age, but we
nevertheless incremented the initial values slightly by
0.005\,$M_{\sun}$ and 0.002\,$M_{\sun}$, respectively, so as to
reproduce the measured masses exactly at the best-fit age.  An
excellent fit to the radii, temperatures, and luminosities of both
stars was found for mixing length parameters of 1.80 for the primary
and 1.50 for the secondary, and convective core overshooting
parameters of 0.35 and 0.30, respectively, with estimated
uncertainties in each of these of about 0.05. Deviations in $R$,
$T_{\rm eff}$, and $L$ from the measured values are all smaller than
0.4$\sigma$. The best-fit age we obtained, about 649 Myr, was
constrained to be the same for the two stars. We illustrate this
solution in Figures~\ref{fig:claret1} and \ref{fig:claret2} for the
$L$--$T_{\rm eff}$ and $R$--$T_{\rm eff}$ diagrams, respectively.

We point out that the age in this best-fit solution is driven entirely
by the properties of the secondary, specifically, by its effective
temperature. This is because that star is in such a rapid phase of
evolution that the temperature is predicted to change drastically (by
many times the observational uncertainty) in just 1~Myr.
Figures~\ref{fig:claret1} and \ref{fig:claret2} show, for example,
that between the ages of 648 and 649 Myr the temperature of the
secondary decreases by 1130\,K, cooling by a further 660\,K over the
next million years. The primary, on the other hand, stays at
essentially the same temperature over this time, and only changes its
radius and luminosity, but at a much slower pace (see top insets in
the figures).  Consequently, it does not constrain the age nearly as
much. Because of the rapid evolution of the secondary, the formal
uncertainty in the age that comes from its temperature error is
negligible. A more meaningful uncertainty may be obtained by varying
its mass within allowed limits, which results in an age range of
approximately $\pm 5$~Myr. This does not include possible systematic
errors having to do with the physics in the models. As found above
from the PARSEC models, the primary star is seen to be located in the
clump, on the hot side of the giant loop (end of core helium-burning
phase), where the radius and luminosity are increasing with time.

A test with Granada models that include rotation for both components
gave a very similar fit, with an age of 655 Myr that is only
marginally older than before.

\begin{figure}
\epsscale{1.15}
\plotone{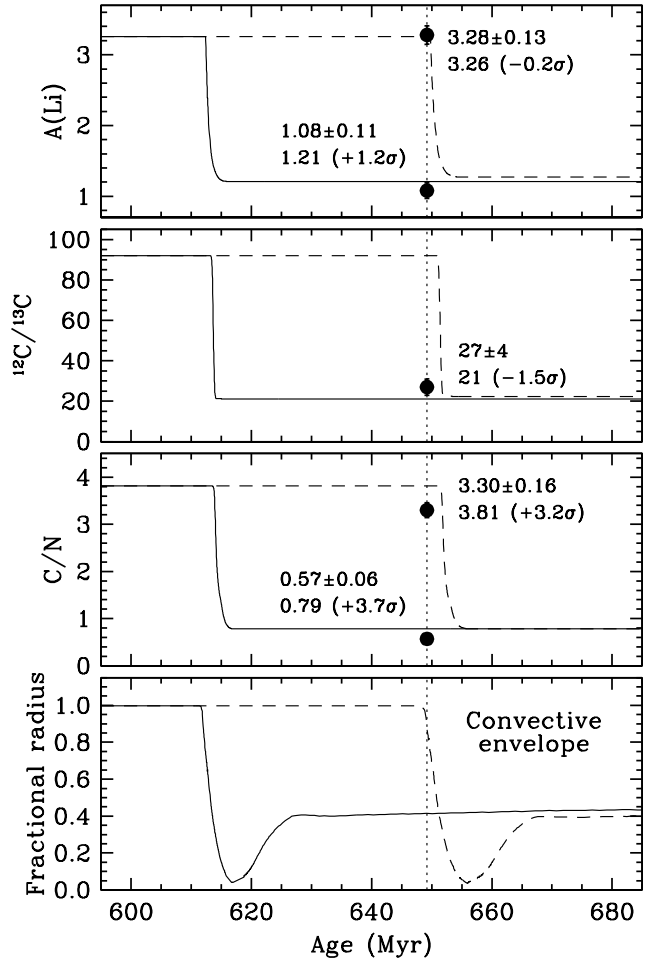}

\figcaption[]{Evolution of chemical diagnostics for Capella occurring
  near the first dredge-up, as predicted by the models of
  \cite{Claret:04}. Solid lines represent the primary, and dashed
  lines the secondary. The best-fit age of about 649 Myr is marked by
  the vertical dotted lines. The lower panel shows the depth of the
  bottom of the convective zone as a function of time, in units of the
  stellar radius. The dots represent the measurements, with error bars
  that are barely visible on this scale. Measured and predicted values
  are listed near each point, with the deviations listed in
  parenthesis next to the predictions, in units of the observational
  errors.\label{fig:dredgeup}}

\end{figure}

The evolution of the surface chemistry in giants is the result of
mixing and is directly related to the depth of the convection zone,
which changes drastically as the stars approach the so-called first
dredge-up, during their initial ascent of the giant branch. Other
changes can occur later. The first dredge-up event is illustrated in
Figure~\ref{fig:dredgeup}, which shows predictions from the Granada
models for the abundance of lithium, the $^{12}$C/$^{13}$C carbon
isotope ratio, and the C/N ratio as a function of time. Also shown for
reference are the changes in the location of the bottom of the
convection zone for each star (lower panel). The measurements of
these key chemical diagnostics are represented with dots at the
best-fit age of 649 Myr. Generally there is good agreement between
theory and observation, except for the C/N ratios that deviate the
most.\footnote{The predicted \emph{difference} between the ratios,
  however, is in better agreement with the measured difference (to
  within 1.7$\sigma$).} We note, however, that the C/N measurements
rely on emission fluxes from spectral lines in the lower transition
layers between the stellar chromosphere and the corona (specifically,
\ion{C}{4}~$\lambda$1550.8 and \ion{N}{5}~$\lambda$1238.8; see T09),
so they may not strictly represent the abundances in the photosphere
\citep[despite some evidence that they do; see,
  e.g.,][]{Bohm-Vitense:92}.  The convection zone in the secondary
star is seen to have just begun deepening, and should reach maximum
depth approximately 7 Myr from now, according to these models.

A final test was performed against stellar evolution tracks computed
using the Modules for Experiments in Stellar Astrophysics \citep[MESA,
  revision 7385; see][]{Paxton:11,
  Paxton:13}.\footnote{http://mesa.sourceforge.net/~.} These models
use the scaled solar abundances of \cite{Asplund:09}, according to
which $Z_{\sun} = 0.0134$. The measured composition of Capella
corresponds to a metal mass fraction of about $Z = 0.012$, and the
helium abundance follows from an adopted $\Delta Y/\Delta Z = 1.67$,
and is $Y = 0.270$. Mass loss is again computed according to the
\cite{Reimers:75} prescription, in this case with an efficiency
parameter (multiplicative scale factor) of $\eta_{\rm R} = 0.2$.  For
this paper we used the `{\tt grey\_and\_kap}' surface boundary
condition \citep{Paxton:13}, with opacities and equation of state as
discussed extensively by \citet{Paxton:11, Paxton:13}.

Overshoot mixing across convective boundaries is treated slightly
differently than in the models considered previously. MESA uses the
exponential decay formalism of \citet{Herwig:97}, in which the product
of the free parameter $f_{\rm ov}$ and the local pressure scale height
provide the scale length over which the diffusion coefficient decays
from its value in the convective region.  Although MESA allows for the
free parameter to take on different values depending on the nuclear
burning present in each convective zone, we have opted to use the same
$f_{\rm ov}$ value for H- and He-burning regions.  For reference, with
the choices listed above, we obtained a solar-calibrated mixing length
parameter $\alpha_{\rm MLT} = 1.84$, and \citet{Herwig:97} suggest a
value of $f_{\rm ov} \simeq 0.02$ for overshoot mixing above the
H-burning core.

\begin{figure}
\epsscale{1.15}
\plotone{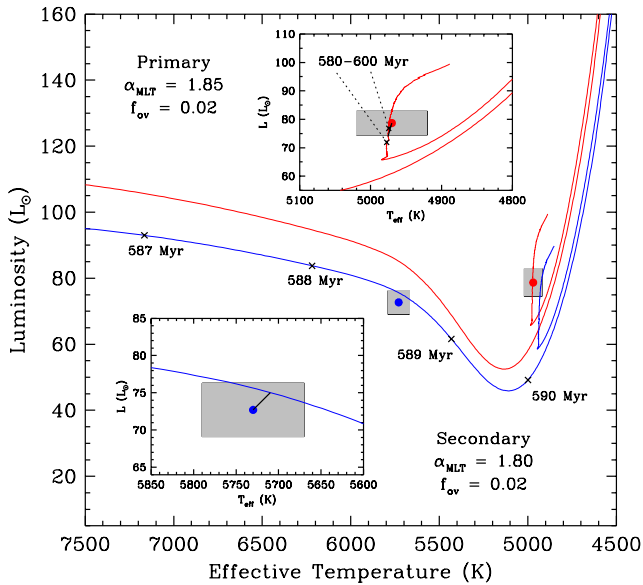}

\figcaption[]{Similar to Figure~\ref{fig:claret1} for the MESA
  models. The measured metallicity corresponds to $Z = 0.0120$ in these
  models.\label{fig:mesa1}}

\end{figure}

\begin{figure}
\epsscale{1.15}
\plotone{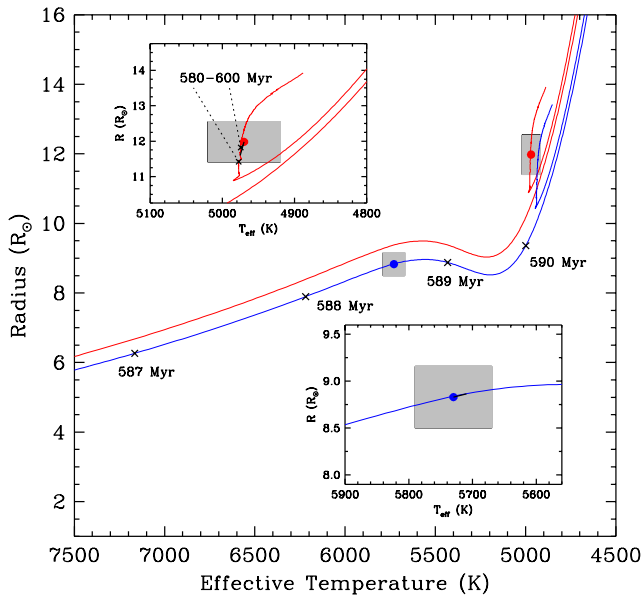}

\figcaption[]{Similar to Figure~\ref{fig:mesa1} for the $R$--$T_{\rm
    eff}$ plane.\label{fig:mesa2}}

\end{figure}

A grid of MESA evolutionary tracks with the specified composition was
computed for each star over broad ranges in $\alpha_{\rm MLT}$ (1.70
to 2.00, in steps of 0.05) and $f_{\rm ov}$ (0.00--0.04, in steps of
0.01). The models were evolved from the fully-convective pre-main
sequence to the end of core He burning. Due to the inclusion of mass
loss, we increased the initial masses of the stars, as we did before
for the Granada models, by 0.0048\,$M_{\sun}$ and 0.0027\,$M_{\sun}$
in this case such that the tracks reach the measured masses at their
respective present locations in the H-R diagram. An excellent
fit to the properties of both components was achieved at a common age
of 588.5~Myr, with all residuals being smaller than 1.2$\sigma$. The
optimal convective parameters were found to be $\alpha_{\rm MLT} =
1.85$ for the primary and $\alpha_{\rm MLT} = 1.80$ for the secondary,
with $f_{\rm ov} = 0.02$ for both stars. The matches in the
$L$--$T_{\rm eff}$ and $R$--$T_{\rm eff}$ diagrams are illustrated in
Figures~\ref{fig:mesa1} and \ref{fig:mesa2}. Once again the models
place the primary at the end of the core helium-burning phase. The age
is somewhat younger than obtained from the Granada models (a
$\sim$10\% difference), while the age found earlier from the PARSEC
models is intermediate between these two. These age differences
correlate strongly and have to do mostly with the $Z$ value used in
the calculations for each model.  The differences in $Z$ at a fixed
(measured) [Fe/H] value are in turn a consequence of the different
heavy-element mixtures adopted for the Sun in each case.

The predictions from the MESA models regarding the evolution of the
surface chemistry of the stars are similar to those from the Granada
models. In particular, the predicted $^{12}$C/$^{13}$C ratio for the
primary at the age of 588.5~Myr is 20.7, and the expected C/N ratios
for the primary and secondary are 0.77 and 3.98, respectively, both
being somewhat higher than measured.

\section{Tidal evolution}
\label{sec:tidalevolution}

The considerable amount of information available for Capella offers a
valuable opportunity to test our understanding of tidal theory in
binary stars, in ways those models have not often been challenged
before. Such calculations are capable of making detailed predictions
about the evolution of the size and shape of the orbit, as well as the
rotational properties of the individual components including their
spin rate, $\Omega_{\rm rot} = 2\pi/P_{\rm rot}$, and the spin-orbit
angle $\phi$ (angle between the spin axis and the total angular
momentum vector of the orbit, sometimes referred to as `obliquity').
In addition to the known orbital elements of Capella, estimates are
available also of the rotation periods of both stars, $P_{\rm rot}$
(see Table~\ref{tab:dimensions}, and T09), and of their projected
rotational velocities, $v \sin i$. Our earlier study of the binary
examined its tidal evolution considering the turbulent dissipation and
radiative damping mechanisms by \citet[and references
  therein]{Zahn:92}, as well as the hydrodynamical mechanism of
\citet[and references therein]{Tassoul:97}. These theoretical
formulations involve a number of assumptions and simplifications
discussed by \cite{Zahn:77} and \cite{Hut:81}. In particular, the
equations are linearized around the equilibrium state, and are
strictly valid only for relatively small eccentricities and
near-synchronous rotation, as well as relatively small obliquities.

In this work we have chosen to use the more general equations of tidal
evolution by \cite{Hut:81}, which are valid for arbitrary
eccentricities and rotation rates, although they are still restricted
to relatively low mutual inclination angles $\phi$. We used a
fourth-order Runge-Kutta algorithm to integrate the six coupled
differential equations describing the time-dependent changes in the
orbital semimajor axis ($da/dt$), eccentricity ($de/dt$), angular
rotational velocities of both stars ($d\Omega_{\rm rot,A}/dt$,
$d\Omega_{\rm rot,B}/dt$), and their spin-orbit angles ($d\phi_{\rm
  A}/dt$, $d\phi_{\rm B}/dt$). In what follows we normalize the
rotation rates to the mean orbital rate $\Omega_{\rm orb}$, for
convenience.  The relevant stellar properties that also evolve with
time, such as the radius, were taken at each integration step directly
from the best-fit evolutionary tracks of \cite{Claret:04} discussed in
the previous section. The turbulent dissipation timescale for the
stellar phases with convective envelopes (later stages for Capella)
was adopted from \cite{Zahn:77}, whereas the timescale for earlier
phases with radiative envelopes follows \cite{Claret:97}.  The initial
conditions, which are unknown, were set by trial and error to match
the measured values of the orbital period, the eccentricity, and the
spin rates at the current age as closely as possible.

\begin{figure}
\epsscale{1.15}
\plotone{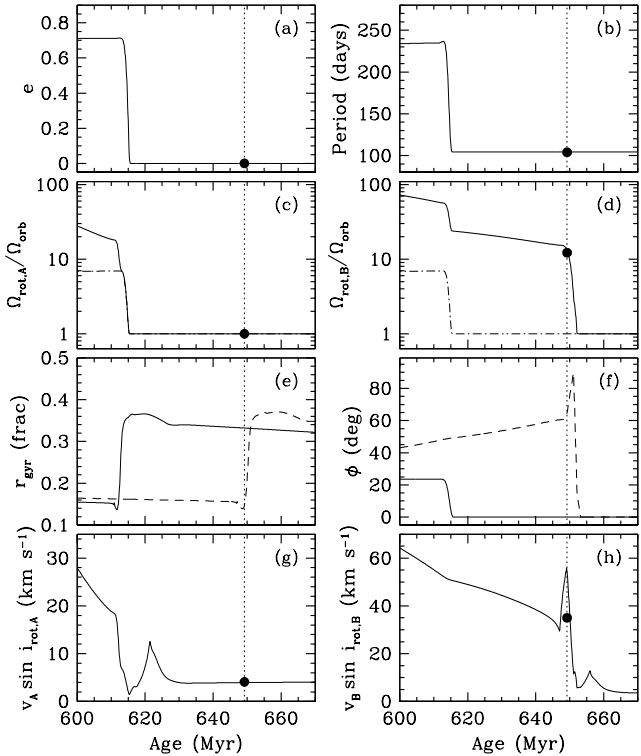}

\figcaption[]{Predicted tidal evolution for Capella according to
  \cite{Hut:81}, compared with the observations (filled circles; error
  bars are smaller than the point size). (a) Eccentricity
  evolution. The vertical dotted line in this and the other panels
  marks the current age of the binary (649 Myr) according to the
  models by \cite{Claret:04} described in the text. (b) Evolution of
  the orbital period. (c) and (d) Evolution of the spin rate of the
  primary and secondary, normalized for convenience in terms of the
  orbital angular velocity. The dot-dashed line represents the
  evolution of the pseudo-synchronous rotation rate, computed
  following Eq.(42) of \cite{Hut:81}. (e) Evolution of the fractional
  gyration radius of each star, for reference (solid line for primary,
  dashed for secondary). (f) Evolution of the spin-orbit inclination
  angle for each star (lines as in previous panel). (g) and (h)
  Predicted projected rotational velocities of the two
  stars.\label{fig:tidal}}

\end{figure}

The outcome of these calculations is illustrated in the top four
panels of Figure~\ref{fig:tidal}.  Setting the initial values to $P_0
= 220$~days, $e_0 = 0.70$, $(\Omega_{\rm rot,A}/\Omega_{\rm orb})_0 =
320$, and $(\Omega_{\rm rot,B}/\Omega_{\rm orb})_0 = 260$ leads to
evolved properties that are very close to those observed at the
present age. In particular, the observed super-synchronous rotation
rate of the secondary ($\Omega_{\rm rot,B}/\Omega_{\rm orb} = P_{\rm
  orb}/P_{\rm rot,B} \approx 12$) is well reproduced.  Qualitatively
the largest difference is perhaps in the orbital eccentricity, which
theory predicts should strictly have fallen to zero some 35 Myr ago,
driven almost exclusively by the evolution of the primary star.
Quantitatively, however, the difference in $e$ between theory and
observation is small, as the measured value (if real) is only $e =
0.00089 \pm 0.00011$. While the agreement reached in these four
observed properties is not entirely unexpected because we have allowed
for four free parameters (the initial values), we note that the good
fit was only possible by increasing the nominal efficiency of the
tidal dissipation by more than an order of magnitude.  Without this
increase, we find that the predicted rotational velocities of the
stars near the zero-age main-sequence are unreasonably low ($v \sin i
\sim 20$\,\kms) for early A-type stars, such as the Capella components
would have been. In order to yield more reasonable projected
rotational velocities in excess of 100\,\kms\ we had to increase the
efficiency of the tidal mechanisms by a factor of $\sim$40. A similar
shortcoming in the efficiency of theory was found earlier by
\cite{Claret:97}, in their analysis of tidal synchronization and
circularization of a sample of detached eclipsing binaries.

Figure~\ref{fig:tidal}f displays the evolution of the spin-orbit
angles for the two stars near the present age, where the integrations
have been performed with arbitrary initial values of $\phi_0 = 0.4$
radians (about 23\arcdeg) for both stars, as we have no direct handle
on those angles. Tests with other (non-zero) values show
that the behavior is qualitatively always the same: the primary's spin
axis aligns itself with the orbital axis (i.e., $\phi$ reaches zero)
well before the current age, whereas the secondary remains formally
misaligned. The alignment of the primary happens at very nearly the
same time that its rotation becomes synchronous and that the orbit
circularizes. The spin-orbit angle of the secondary quickly shrinks to
zero shortly after the present age (4--5 Myr later), at which time it
also synchronizes its rotation with the mean orbital motion. In both
cases the changes are the result of significant structural adjustments
in the stars, such as a sharp increase in the radius of gyration,
$r_{\rm gyr}$, related to the moment of inertia through $I = M (r_{\rm
  gyr} R)^2$ (see Figure~\ref{fig:tidal}e).

\begin{figure}

\epsscale{1.15}
\plotone{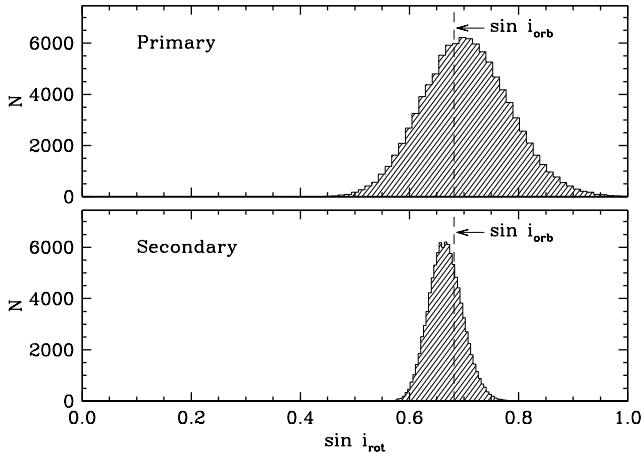}

\figcaption[]{The histograms represent the empirical distributions of
  $\sin i_{\rm rot}$ derived from the measurement of $v \sin i$,
  $P_{\rm rot}$, and the radius of each star as given in
  Table~\ref{tab:dimensions}. The dashed lines mark the value of $\sin
  i_{\rm orb}$ determined from our orbital fit. The good agreement
  strongly suggests that the spin axes of both stars may in fact be
  parallel to the orbital axis.\label{fig:sini}}

\end{figure}

While we cannot measure the present-day values of $\phi$, it is
possible to gain indirect knowledge about these angles using our
spectroscopic estimates of $v \sin i$ along with the measurements of
$P_{\rm rot}$ and the radii of both stars. These quantities are
trivially related by
\begin{equation}
\label{eq:vsini}
v \sin i = \frac{2\pi R}{P_{\rm rot}} \sin i_{\rm rot}\,,
\end{equation}
where the inclination angle on the left-hand side (inaccessible to
direct observation) is strictly that of the stellar spin axis relative
to the line of sight ($i_{\rm rot}$), and also appears on the right.
We may thus solve for the $\sin i_{\rm rot}$ term on the right-hand
side.  Figure~\ref{fig:sini} displays the distributions of $\sin
i_{\rm rot}$ derived from the propagation of all observational errors
in a Monte Carlo fashion. For reference we also show the sine of the
orbital inclination angle relative to the line of sight ($i_{\rm
  orb}$), which is directly measurable from the astrometric
observations (Table~\ref{tab:dimensions}). The close agreement between
$\sin i_{\rm rot}$ and $\sin i_{\rm orb}$ for both stars is highly
suggestive that the spin axes may actually be parallel to the orbital
axis in space, and this in turn would imply $\phi = 0$.  We cannot
rule out a difference in quadrants such that the spin axes are tilted
with respect to the axis of the orbit while still maintaining the same
sine value (e.g., $i_{\rm rot} = 180\arcdeg - i_{\rm orb}$), but such
a coincidence for both stars seems rather unlikely.\footnote{Barring
  this type of situation, conversion of the distributions in
  Figure~\ref{fig:sini} to inclination angles yields $i_{\rm rot,A} =
  135.3^{+6.1}_{-6.8}$ deg and $i_{\rm rot,B} = 138.2^{+2.3}_{-2.4}$
  deg, compared to the orbital value of $i_{\rm orb} = 137.156 \pm
  0.046$ deg.} This indirect empirical evidence that the obliquity may
currently be zero for both stars (which would hardly happen by chance) appears to point to a discrepancy
with the prediction from tidal theory for the secondary component
(Figure~\ref{fig:tidal}f), whereas for the primary there is good
agreement.

It is possible that part of the reason for this difference lies in the
small-angle approximation implicit in the tidal differential equations
of \cite{Hut:81} for the angle $\phi$. However, we note also that the
secondary of Capella happens to be in such a rapid state of evolution
that theoretical predictions for the spin-orbit angle are extremely
sensitive to other details, including those related to structural
changes in the stars happening at this stage. Those changes may actually have
a larger impact than the dissipation processes themselves (though the latter must
also matter). In particular, changes in both the spin rate and the
spin-orbit angle are strongly driven in part by the sudden change in
the moment of inertia (or equivalently, the gyration radius)
illustrated in Figure~\ref{fig:tidal}e. An additional complication is
the large ad-hoc increase in the efficiency of the tidal mechanism
that was required to match other observations, as mentioned earlier.
Thus, a definitive assessment of the accuracy of tidal theory related
to its other approximations is difficult in this case.

An alternate way of comparing observations with the predictions from
tidal and evolution models combined is by examining the evolution of
$v \sin i$, which is a spectroscopically measured quantity. It is a
function of the stellar radius, the spin rate, and the inclination
angle of the rotation axis to the line of sight, $i_{\rm rot}$
(Eq.[\ref{eq:vsini}]). The latter angle is not directly predicted by
theory, but is related to other angles by
\begin{equation}
\label{eq:lambda}
\cos\phi = \cos i_{\rm orb} \cos i_{\rm rot} + \sin i_{\rm orb} \sin i_{\rm rot} \cos\lambda\,,
\end{equation}
in which the obliquity $\phi$ can be predicted, $i_{\rm orb}$ is
known, and $\lambda$ is the angle between the sky projections of the
spin axis and the orbital axis (`projected obliquity'). The angle
$\lambda$ changes with time and is directly measurable in eclipsing
systems by observing the Rossiter-McLaughlin effect. As Capella does
not eclipse we have no knowledge of this angle in this case, except
for the weak condition that it represents a lower limit to the
three-dimensional angle $\phi$ \citep[see, e.g.,][]{Fabrycky:09}.
Eq.[\ref{eq:lambda}] may be solved for the quantity $\sin i_{\rm rot}$
that we need in order to compute $v \sin i$ with eq.[\ref{eq:vsini}],
resulting in the quadratic equation
\begin{equation*}
A \sin^2 i_{\rm rot} + B \sin i_{\rm rot} + C = 0
\end{equation*}
where
\begin{align*}
A &= \cos^2 i_{\rm orb} + \sin^2 i_{\rm orb} \cos^2\lambda \\
B &= -2 \cos\phi\, \sin i_{\rm orb} \cos\lambda \\
C &= \cos^2\phi - \cos^2 i_{\rm orb}\,.
\end{align*}
For the primary star the prediction that $\phi = 0$ near the current
epoch means that $\lambda$ is also zero. Using this value in the
equations above to compute the expected evolution of the projected
rotational velocity leads to the trend shown in
Figure~\ref{fig:tidal}g, where we refer to $v \sin i$ more properly as
$v_{\rm A} \sin i_{\rm rot,A}$. Values at times when $\phi \neq 0$
will be somewhat less accurate because the projected obliquity
$\lambda$ may also be different from zero.  The predicted value of
4.0~\kms\ at the current age is in excellent agreement with the
measurement ($4.1 \pm 0.4$~\kms).  The corresponding evolution of
$v_{\rm B} \sin i_{\rm rot,B}$ is seen in Figure~\ref{fig:tidal}h. In
this case the predicted value at the current age does not match the
measurement, and the discrepancy is in part a reflection of the
evolution of $\phi$ that was discussed above, but also has to do with
the very rapid changes in the structure of the star ($R$, $r_{\rm
  gyr}$) at the present time. Tests in which we changed $\lambda$
within reason yielded very similar results, and cannot explain the
difference.

Finally, it is of interest to verify that the components of Capella
have always been detached, as any significant mass transfer earlier in
their lives (e.g., through Roche lobe overflow in the primary) would
invalidate our comparison with stellar evolution models, which are
designed for normal (unperturbed) stars. The fraction of its Roche
lobe filled by the primary component depends on its size and also on
the size of the orbit, which changes due to tidal forces
(Figure~\ref{fig:tidal}b).  The same Granada evolutionary track used
above indicates the star attained a maximum radius at the tip of the
giant branch of about 38\,$R_{\sun}$ at an age of 617~Myr (MESA models
predict a similar maximum size of 36\,$R_{\sun}$), and the filling
factor ($R/R_{\rm Roche}$) at the time was approximately 0.63. This
indicates that mass transfer through Roche-lobe overflow has not taken
place in Capella.

\section{Discussion and concluding remarks}
\label{sec:discussion}

For the first time current stellar evolution models are shown here to
provide a satisfactory fit to the observed global properties of both
components of Capella at a single age, for a chemical composition
equal to that measured.  The comparison confirms the long-held but
largely unsubstantiated belief that the primary is a clump star, and
more precisely, it suggests Capella~A is near the end of its core
He-burning phase, as indicated consistently by three different sets of
models. The principal factors that have allowed the better match are
an improvement in the accuracy of the absolute masses, made possible
by the high-quality radial-velocity measurements of \cite{Weber:11},
and the first robust determination of the metallicity of Capella
derived here. Our detailed chemical analysis of the disentangled
spectra of the components yields essentially the same near-solar
composition for the two stars, which is rather different from the
sub-solar abundance the binary was previously assumed to have in our
earlier T09 study. All measured properties (masses, radii,
temperatures, and independently derived luminosities) are now
simultaneously in agreement with the models to within
0.4--1.4$\sigma$, depending on the model.  This result strengthens our
confidence in evolutionary calculations for evolved stars.

Chemical indicators of evolution that differ greatly between the
components, such as the $^{12}$C/$^{13}$C ratio, the lithium
abundance, and the C/N ratio, are also broadly in agreement with
theoretical predictions, though with somewhat larger differences that
may be due to either shortcomings in the mixing prescriptions in the
models, or observational errors in these delicate measurements. Rarely
has it been possible to perform this type of test involving key
chemical diagnostics for a binary with two evolved components having
well-determined absolute dimensions ($M$, $R$, etc.), such as Capella.

On the other hand, the performance of tidal theory as described by the
tidal evolution equations of \cite{Hut:81} is not as good.  As was
noticed previously by \cite{Claret:97}, we again find that the
efficiency of the tidal mechanisms involved seems to be too low by a
factor of $\sim$40, if predictions are to be consistent with the large
rotational velocities Capella A and B presumably had earlier in their
lives as A-type stars ($v \sin i > 100$\,\kms), and at the same time
the much lower velocities they now have as evolved stars.  Application
of an ad-hoc adjustment of this magnitude brings agreement, although
theory still forecasts a significant misalignment between the spin
axis of the secondary and the orbital axis at the present age, whereas
empirical evidence based on the direct measurement of the star's
radius, rotation period, and $v \sin i$ seems to favor an obliquity
consistent with zero.  Model predictions about the synchronous
rotation of the primary and its spin-orbit alignment do agree with
observational clues, as does the supersynchronous rotation of the
secondary. A small additional disagreement in that the orbit seems to
be very slightly eccentric even though it is expected to be perfectly
circular may or may not be significant, as the difference is small.

Finally, another important consequence of our revised chemical
composition for Capella concerns the relationship with the coronal
abundances, which have been measured by many authors as summarized by
T09. Coronal abundances in the Sun are known to depend on the first
ionization potential (FIP), in such a way that low-FIP elements (less
than 10 eV) are overabundant compared to those with higher FIP
\citep[see, e.g.,][]{Feldman:02}. T09 showed this to be the case for
Capella as well.  Other stars display the opposite effect \citep[see,
  e.g.,][]{Brinkman:01, Laming:04}. What is less clear from evidence
in other stars is whether the low-FIP elements are enhanced relative
to the photospheric abundance, or whether it is the high-FIP elements
that are depleted compared to the photosphere. T09 adopted a sub-solar
photospheric composition for Capella of ${\rm [m/H]} = -0.34 \pm
0.07$, based on the work of \cite{McWilliam:90}, and since this agreed
with the measured coronal abundances of the high-FIP elements (see
their Figure~18), they concluded it was the low-FIP elements that were
enhanced.  Interestingly, our revised photospheric abundance of ${\rm
  [Fe/H]} = -0.04 \pm 0.06$ leads to precisely the opposite
conclusion: now it is the low-FIP elements that agree with the
photosphere, and therefore the high-FIP elements are depleted. A
recent study by \cite{Peretz:15} of the coronae of stars with
super-solar photospheric abundances found the same effect shown by
Capella in $\alpha$~Cen A and B, which have somewhat similar
temperatures, although these are dwarfs so it is unclear how
significant this may be. In any case, Capella now represents a robust
point of reference for such studies in evolved and active stars.

\acknowledgements

We thank Brian Mason (U.S.\ Naval Observatory) for providing the
historical positional measurements of the Capella HL system, and the
anonymous referee for helpful comments.  The research has made use of
the SIMBAD database, operated at CDS, Strasbourg, France, of NASA's
Astrophysics Data System Abstract Service, of the Washington Double
Star Catalog maintained at the U.S.\ Naval Observatory, and of data
products from the Two Micron All Sky Survey (2MASS), which is a joint
project of the University of Massachusetts and the Infrared Processing
and Analysis Center/California Institute of Technology, funded by NASA
and the NSF.

\appendix

\section{Capella as a member of a multiple system}
\label{sec:capellaHL}

Over the last two centuries at least half a dozen visual companions to
Capella have been recorded at angular separations ranging from
47\arcsec\ to 12\arcmin. Only the widest one, discovered by
\cite{Furuhjelm:14}, has been shown to be physically associated, and
has received the double-star designation Capella~H (also ADS~3841\,H,
GJ~195\,A, and LTT~11622, among others). The projected linear
separation is $\sim$9500~au (723\arcsec\ in position angle
141\arcdeg). This 10th-magnitude star is of spectral type M1--M2.5,
depending on the source, and has the same relatively large proper
motion as Capella itself \citep{Furuhjelm:14, Lepine:07} as well as
the same parallax \citep[e.g.,][]{Daniel:20, Adams:26,
  Bagnuolo:89}. The earliest reported radial velocity measurements
appear to be those of \cite{Abt:70} at Mount Wilson, of which the
first two from consecutive nights in 1921 gave a mean of about
$+30$~\kms, while a third from 1927 is $+48.5$~\kms, with a large
probable error. \cite{Stauffer:86} reported a mean radial velocity of
$+30 \pm 1$~\kms\ based on an unspecified number of CfA spectra using
the same setup as T09, and \cite{Upgren:88} listed three measurements
also from CfA taken in 1986, giving similar values.  The first two of
these are likely to be based on the same spectra used by
\cite{Stauffer:86}.  Other measurements were reported by
\cite{Reid:95} ($+31.9 \pm 15$~\kms) and \cite{Gizis:02} ($+32.7 \pm
1.5$~\kms).

In order to supplement these values we report eight additional
observations of Capella~H with the same instrumentation and procedures
described by T09.  Radial velocities were derived with the
IRAF\footnote{IRAF is distributed by the National Optical Astronomy
  Observatories, which is operated by the Association of Universities
  for Research in Astronomy, Inc., under contract with the National
  Science Foundation.} task {\tt xcsao}, using a synthetic template
selected to match the properties of the star with $T_{\rm eff} =
3750$~K and no rotational broadening. The three spectra obtained by
\cite{Stauffer:86} and \cite{Upgren:88} were reanalyzed in the same
way, and our velocities supersede those in the original papers. We
list all these measurements in Table~\ref{tab:capellah}, on the
absolute heliocentric frame defined by the IAU Radial-Velocity
Standard Stars \citep[see][]{Stefanik:99}.  The 11 velocities show no
variation within the measurement errors, and have a mean of $+31.63
\pm 0.14$~\kms. This is $1.69 \pm 0.14$~\kms\ larger than the
center-of-mass velocity of Capella reported in
Table~\ref{tab:elements}, and is most likely due to motion in the
$\sim$9500~au orbit.\footnote{Part of the difference ($\sim$0.5~\kms)
  is due to the larger gravitational redshift of the M dwarf compared
  to the pair of giants, and there could also be a contribution from
  the difference in convective blueshifts. Additionally, there may be
  a slight offset between the zero-point of the \cite{Weber:11}
  velocities and the IAU system.}

\begin{deluxetable}{lccc}
\tablewidth{0pc}
\tablecaption{New radial velocity measurements of Capella~H.\label{tab:capellah}}
\tablehead{
\colhead{HJD} &
\colhead{} &
\colhead{RV} &
\colhead{$\sigma_{\rm RV}$} \\
\colhead{\hbox{~~(2,400,000$+$)~~}} &
\colhead{Julian Year} &
\colhead{(\kms)} &
\colhead{(\kms)}
}
\startdata
    46475.6230\dotfill &   1986.121 &   +31.33 &     0.82 \\
    46494.5453\dotfill &   1986.173 &   +30.97 &     0.82 \\
    46788.8338\dotfill &   1986.978 &   +31.84 &     0.70 \\
    52986.8030\dotfill &   2003.947 &   +32.15 &     0.63 \\
    53030.7455\dotfill &   2004.068 &   +32.19 &     0.65 \\
    53055.6708\dotfill &   2004.136 &   +32.16 &     0.55 \\
    53078.6029\dotfill &   2004.199 &   +31.71 &     0.51 \\
    53106.5530\dotfill &   2004.275 &   +31.43 &     0.67 \\
    53361.7995\dotfill &   2004.974 &   +31.93 &     0.58 \\
    53419.6511\dotfill &   2005.133 &   +31.09 &     0.58 \\
    53689.8944\dotfill &   2005.872 &   +31.08 &     0.56 
\enddata
\end{deluxetable}

\begin{figure*}
\epsscale{0.6}
\plotone{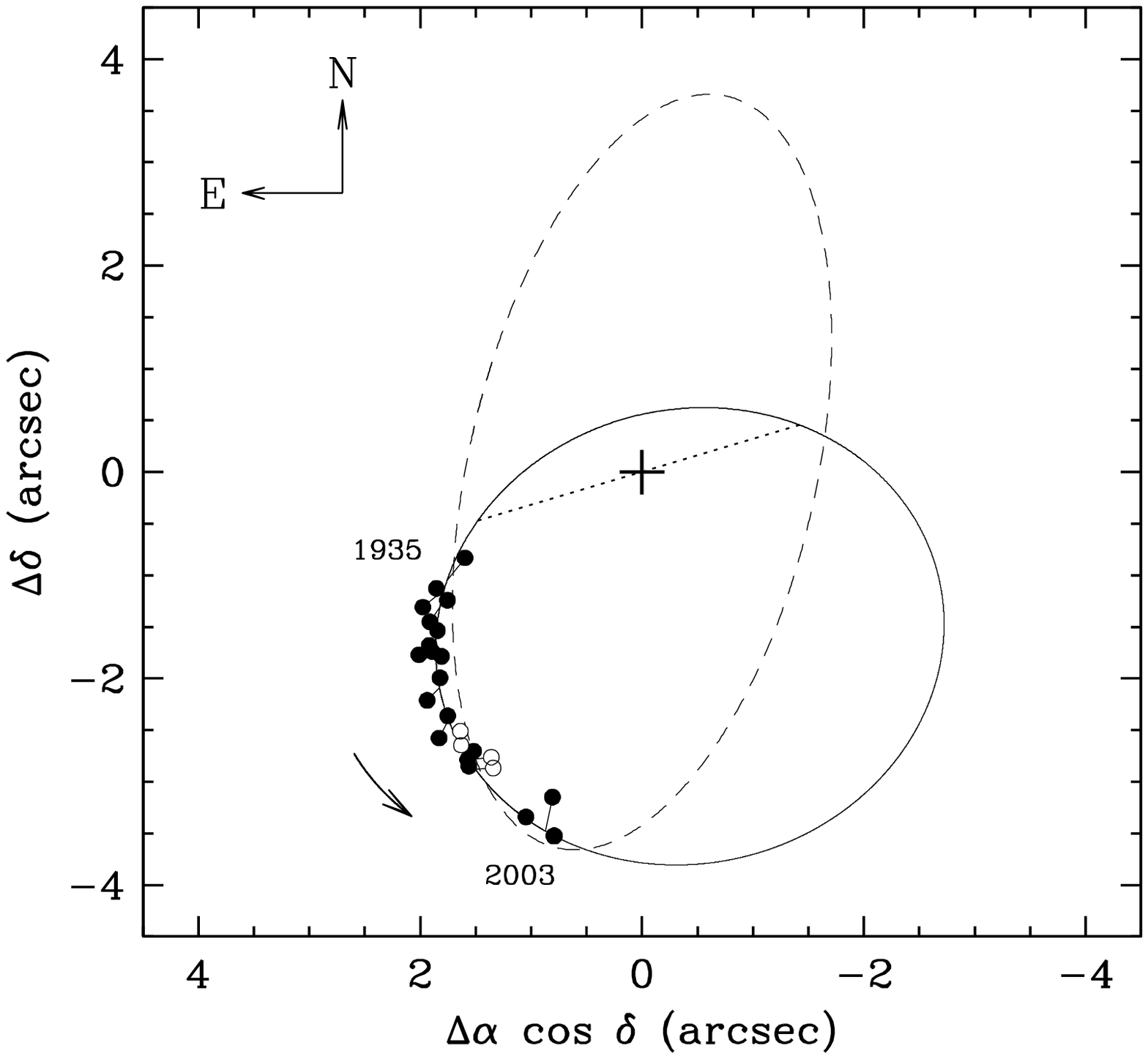}

\figcaption[]{Measurements of the relative position of the Capella~HL
  binary, along with the 388-yr orbit by \cite{Heintz:75} (dashed),
  and our revised orbit (solid). Filled symbols represent measures
  from the Washington Double Star Catalog, \cite{Stearns:39}, and
  \cite{Helminiak:09}, and open symbols are for the photographic
  measurements by Heintz. The line of nodes in the revised orbit is
  indicated with a dotted line, and the motion on the plane of the sky
  is direct (arrow). The primary (Capella H) is represented by the
  plus sign. Short line segments connect the measurements with their
  predicted location on the revised orbit.\label{fig:capellahl}}

\end{figure*}

Capella~H was found by \cite{Stearns:36} to have a close
($\sim$2\arcsec) companion approximately 3.5 magnitudes
fainter\footnote{This visual brightness estimate by \cite{Kuiper:36}
  may be overestimated, however, since the 2MASS $JHK_s$ magnitudes of
  the individual components differ by only a few tenths of a
  magnitude. Alternatively, the companion may itself be multiple.},
designated Capella~L (also ADS~3841\,L and GJ~195\,B). Significant
changes in position relative to Capella~H since the discovery date
indicate orbital motion. Thus, the Capella system is a hierarchical
quadruple.  Historical measurements of the Capella~HL binary from the
Washington Double Star Catalog were kindly provided by Brian Mason,
and are shown in Figure~\ref{fig:capellahl}. They were obtained mostly
with visual micrometers or photographically. We have supplemented
these observations with additional photographic measurements reported
by \cite{Stearns:39} and \cite{Heintz:75}, as well as more recent and
accurate adaptive optics measurements by \cite{Helminiak:09}. A
tentative 388-yr orbit by \cite{Heintz:75} based on his photographic
data and earlier observations is also shown, but does not seem to
represent the bulk of the observations very well.

While the astrometric orbit is still largely undetermined because of
the short arc covered by the measurements, a helpful constraint may be
obtained by making use of rough estimates of the masses of the stars
derived from existing near-infrared photometry (2MASS) and the
empirical mass-luminosity relation of \cite{Delfosse:00}. With
inferred values of approximately 0.57\,$M_{\sun}$ and 0.53\,$M_{\sun}$
for Capella H and L, and the use of the orbital parallax from
Table~\ref{tab:elements}, Kepler's Third Law then leads to a
constraint on the ratio $a^3/P^2$. A tentative new orbital solution
derived in this way, which is also shown in
Figure~\ref{fig:capellahl}, provides a significantly better fit with
standard visual elements $P_{\rm orb} \approx 300$~yr, $a \approx
3\farcs5$, $e \approx 0.75$, $i \approx 52\arcdeg$, $\omega_{\rm L}
\approx 88\arcdeg$, $\Omega_{\rm J2000} \approx 288\arcdeg$, and
periastron passage near the year 2220.

\end{document}